\documentclass{iopjournal}
\pdfoutput=1
\usepackage[utf8]{inputenc}
\usepackage[english]{babel}
\usepackage[T1]{fontenc}
\usepackage{amsmath}
\usepackage{hyperref}
\usepackage{tikz}
\usepackage{lipsum}
\usepackage[numbers]{natbib}

\usepackage{xcolor}
\usepackage{hyperref}
\hypersetup{
    colorlinks=true,
    linkcolor=blue,
    citecolor=blue,
    urlcolor=blue
}

\usepackage{graphicx}
\usepackage{amsfonts}
\usepackage{amssymb}
\usepackage{amsmath}
\usepackage{bbold}
\usepackage{color}
\usepackage{dsfont}
\usepackage{multirow}
\usepackage{appendix}
\usepackage{mathrsfs}
\usepackage{latexsym}
\usepackage{physics}
\usepackage{epigraph}
\usepackage{comment}
\usepackage{tikz-cd}

\usetikzlibrary{decorations.pathmorphing, positioning}
\usetikzlibrary{arrows.meta, decorations.pathmorphing, calc}

\newcommand{\msf}[1]{\mathsf{#1}}
\newcommand{\mtt}[1]{\mathtt{#1}}

\newcommand{\UU}{{\mtt{U}}}
\newcommand{\VV}{{\mtt{V}}}

\definecolor{mcrcolor}{rgb}{0.7,0.2,0.7}

\newcommand{\mcrc}[1]{{\color{mcrcolor}{[MCR: #1]}}}

\begin{document}
\fancyhead[RO]{M. Clavero-Rubio {\it et al.}}
\title{\bf{Topologically protected long-range correlations in steady states of driven-dissipative bosonic chains}}

\author{%
\bf{Miguel Clavero-Rubio}$^{1,2}$\orcid{0009-0002-7132-7097}, 
\bf{Tomás Ramos}$^{1,2}$\orcid{0000-0003-2182-7878}, 
\bf{ Diego Porras}$^{1,2}$\orcid{0000-0003-2995-0299
}}

\affil{$^{1}$Quantum Advanced Research Center (QuARC), CSIC, Calle Serrano 113b, 28006 Madrid, Spain}
\\
\affil{$^{2}$Institute of Fundamental Physics (IFF), CSIC, Calle Serrano 113b, 28006 Madrid, Spain}

\email{miguel.clavero@iff.csic.es, diego.porras@csic.es}


\begin{abstract}
Driven–dissipative quantum systems can exhibit robust transport and amplification in topological regimes, yet the connection between topology and the extent of correlations remains largely unexplored. In this work, we develop a general framework that links topological phases in driven–dissipative systems to bosonic correlations via the singular value decomposition (SVD). In essence, we claim that non-Hermitian topology in quadratic Liouvillians is directly encoded in steady-state correlations, providing an intrinsic characterization of topology without external probes.
 We show that topological amplification induces disorder-robust long-range order (LRO) in steady-state correlations at fixed frequency, establishing frequency-resolved correlations as direct signatures of non-Hermitian topological phases. We introduce a vector-valued topological invariant that captures the total number of singular-value gap closings across the frequency axis, extending the concept of adiabatic deformation
from topological insulators to the case of topological phases of quadratic
Liouvillians. 
Within this framework, we further demonstrate that the spatial structure of equal-time correlations encodes global topological information, manifested as a Gaussian spatial decay with distance in the topological phase, in contrast to the exponential decay characteristic of trivial phases. These findings open new avenues for quantum sensing and correlation engineering in topological non-Hermitian systems, with feasible implementations in platforms such as trapped ions and superconducting circuits.
\end{abstract}

\newpage

\section{Introduction}
The interplay between physical systems and topology has profoundly transformed our understanding of phase transitions by replacing conventional local order parameters with robust global topological invariants. Since its emergence in condensed matter systems, most notably through the discovery of the quantum Hall effect \cite{klitzing1980new} and the subsequent identification of topological insulators \cite{hasan2010colloquium,qi2011topological}, the concept of topological phases has been extensively explored. 
This sustained interest is largely driven by applications of topology in quantum technologies, including quantum computing \cite{kitaev2003fault}, quantum sensing \cite{koch2022quantum,slim2024optomechanical,clavero2025vibrational}, quantum amplification \cite{porras2019topological,wanjura2020topological}, and frequency conversion \cite{parra2025floquet}.

Driven–dissipative bosonic lattices are a natural arena for exploring topological phases beyond equilibrium \cite{bardyn2013topology,pernet2022gap,villa2025topological}. 
In these systems, topology can arise from the interplay between coherent processes and dissipation. 
In the non-interacting case, dissipative bosonic lattices are described by master equations with quadratic Liouvillians that admit exact numerical or even analytical solutions. 
Gain and loss processes lead to two kinds of Liouvillian terms: 
(i) quantum jump terms, and (ii) non-jump terms that govern the system’s evolution in the intervals between quantum jumps. 
The latter can be described by an effective non-Hermitian Hamiltonian, whose topological features have been extensively explored within the framework of non-Hermitian topology \cite{kawabata2019symmetry,ashida2020non,okuma2023non}. 
Non-Hermitian matrices exhibit a range of phenomena that have no direct Hermitian counterparts, like  
the skin effect \cite{okuma2020topological}.
Moreover, the complex energy spectrum intrinsic to non-Hermitian dynamics enables the definition of point-gap topology \cite{yokomizo2019non}, a topological invariant that has no analog in closed Hermitian systems. 

Topological effects in driven-dissipative bosonic lattices require the full quantum solution of the master equation, including both quantum jump and non-jump processes. One of the most striking features in these systems is topological amplification, which refers to the robust, nonreciprocal transport of signals accompanied by an exponential enhancement of their amplitude, while remaining immune to backscattering \cite{porras2019topological,wanjura2020topological,peano2016topological,brunelli2023restoration,gomez2022bridging,gomez2023driven}. 
To date, most theoretical and experimental efforts have focused on identifying the underlying topological phases via the system's linear response to external driving fields, particularly by analyzing the directional amplification of coherent input signals. 
This connection is well understood through the SVD of the non-Hermitian dynamical matrix \cite{porras2019topological,herviou2019defining}, where the emergence of topological singular values underpins the exponential gain characteristic of topological amplification. This approach is closely related to the concept of non-Hermitian point-gap topology. However, a central open question in non-Hermitian quantum matter is how to diagnose topology without relying on conventional approaches based on spectral or dynamical signatures.

\begin{figure}[t]
    \centering
    \includegraphics[width=0.75\textwidth]{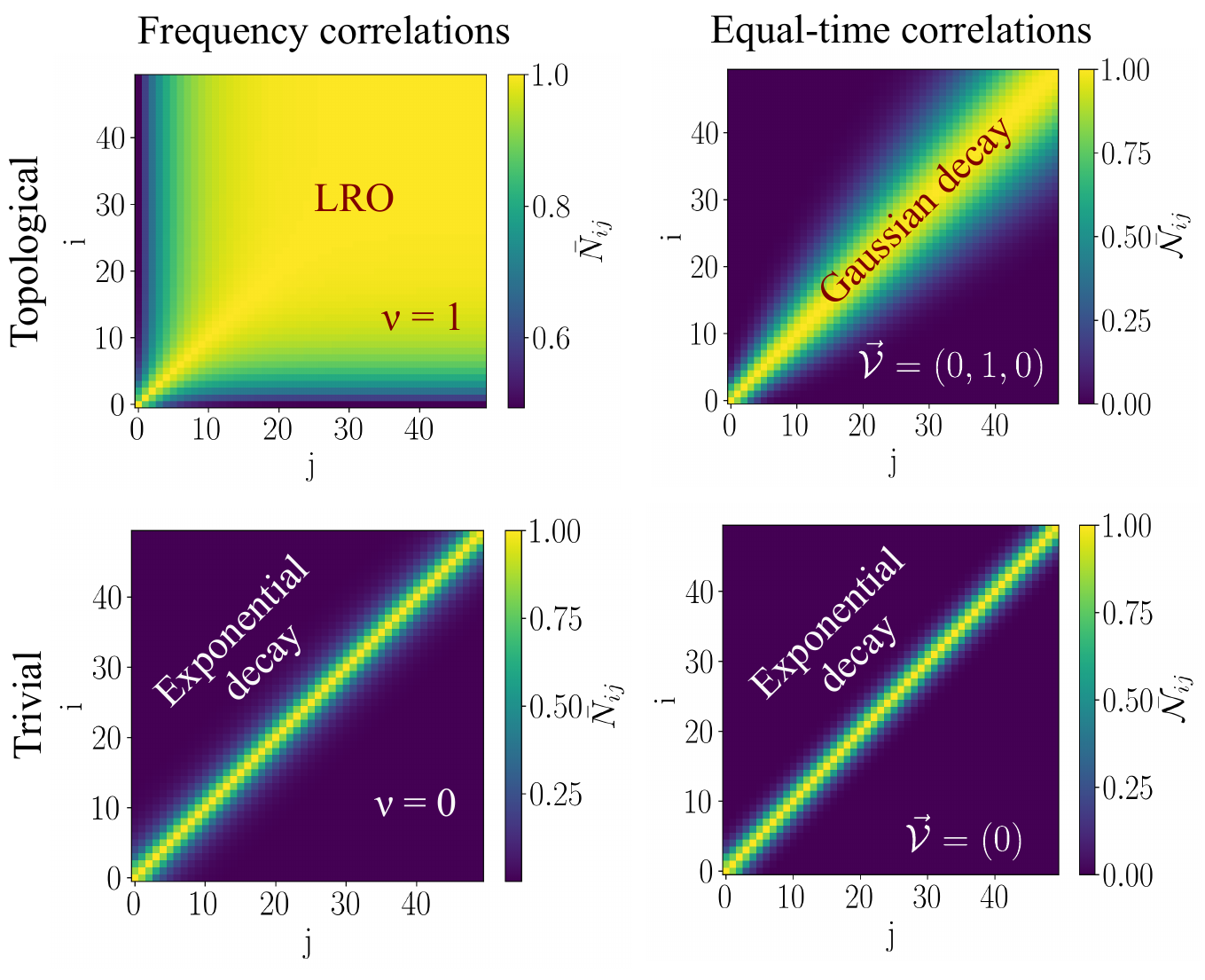}
    \caption{Schematic representation of normalized frequency-resolved ($\bar{N}_{ij}$) and equal-time ($\bar{\mathcal{N}}_{ij}$) correlations in topological and trivial regimes. In the topological phase, frequency-resolved correlations exhibit long-range order, while equal-time correlations display a Gaussian decay. In the trivial regime, both correlations decay exponentially with distance. These results establish two-point correlations as a key steady-state observable for diagnosing non-Hermitian topology. The frequency-resolved topological invariant is denoted by $\nu$, while steady-state topology is characterized by the winding-number array $\Vec{\mathcal{V}}$. Both observables and invariants are defined below.}
    \label{Fig1}
\end{figure}

In this work, we investigate the steady-state observables of topological quadratic Liouvillians, a scenario that has recently attracted significant interest \cite{bardyn2013topology, Flynn21prl, Flynn23prb,
okuma2026steadystateskineffectbosonic,gu2026nonreciprocityenrichedsteadyphasesopen}. 
Rather than relying on response-based probes, we introduce a novel set of intrinsic observables constructed from two-point correlation functions in the steady state. In particular, correlations emerge as a powerful alternative since they are intrinsically defined, experimentally measurable, and directly encode the structure of amplification and mode selection, thereby offering a fundamentally new route to access topological properties.
Previous theoretical works have focused on the buildup of large boson numbers in topological steady-states \cite{gomez2023driven,ramos2021topological,mcdonald2022prb}, and they have shown that non-Hermitian topology manifests in steady-state properties, although the inclusion of quantum jump terms is essential for a complete characterization of the system. 
In our present work, we build on the description of topological amplification in the SVD formalism \cite{porras2019topological,ramos2021topological,gomez2023driven}, and focus on steady-state correlations in absence of external signals. This provides a new paradigm to detect topology without external probes. Our main results are the following:

\begin{itemize}

\item By using the SVD formalism, we extend the concept of adiabatic deformation from topological insulators to the case of topological phases of quadratic Liouvillians. This approach yields a definition of a topological invariant in the form of a winding-number array, which characterizes topologically equivalent quadratic Liouvillians.
\item We show that two-point correlations exhibit a simple spatial structure in topological steady states, typically governed by a single topological singular vector. This result leads to the emergence of topologically protected long-range order in normalized frequency-resolved correlations, as observed in the upper-left plot of Fig. \ref{Fig1}. 
\item We numerically investigate the effect of disorder and show that two-point frequency-resolved correlations remain robust, as a direct consequence of topological protection.
\item We numerically study two-point equal-time correlation functions and show that normalized correlations present an approximate Gaussian decay in the topological regime (as opposed to exponential decay in stable non-topological regimes \cite{ughrelidze2026quantumcriticalitythermodynamicstability}). Such Gaussian spatial decay arises from interference in the convolution of correlation functions at different frequencies. This can be observed in the upper right plot of Fig. \ref{Fig1}.
\item As a side result, we have carried out an analytical calculation of topological singular vectors and values that allowed 
us to obtain the exact form of correlation functions analytically in highly symmetric regimes.
\end{itemize}

Our results go beyond model-specific features and reveal a general principle: topological phases in non-Hermitian quadratic systems can be characterized by the extent of the steady-state correlations. This phenomenon emerges from the presence of a dominant singular value, whose existence is protected by a topological invariant. This establishes a direct connection between singular-value topology and observable correlation functions. We support this picture analytically and numerically in two paradigmatic models and discuss its robustness and physical implications.

The article is organized as follows. In Sec.~\ref{sec:2}, we introduce the general master equation for quadratic Liouvillians, with particular emphasis on steady state correlations in both frequency and time domains. In Sec.~\ref{section2}, we present the formalism employed throughout the manuscript, establishing the connection between the Green’s function, expressed in the singular-value basis, and the bosonic frequency-resolved and equal-time correlations. Section~\ref{section3} is devoted to the driven-dissipative models under consideration. We introduce two models: a homogeneous BdG chain and a dimerized variant, the latter enabling collective gain upon elimination of the auxiliary sites. In Sec.~\ref{section4}, we present the topological characterization of steady state observables of quadratic Liouvillians by introducing the winding-number array, which captures the number of singular value gap closings across the full frequency spectrum. In Sec.~\ref{section5}, we present the main results of the article, demonstrating the emergence of disorder-robust LRO in the frequency-resolved correlations within the topological phase. In contrast, equal-time correlations do not exhibit LRO, although they still display an enhanced spatial extent manifested in Gaussian decay. Finally, conclusions and outlook are presented in Sec.~\ref{section9}.

\section{Driven-dissipative parametric chains}\label{sec:2}
We investigate the steady state properties of quadratic dissipative chains with $N$ bosonic modes, described by the  annihilation and creation operators $a_j$ and $a_j^\dagger$. The dynamics is described by the following Lindblad master equation (units such that $\hbar = 1$)
\begin{equation}
\dot{\rho} = - i [ H, \rho ] + {\cal L}_{\rm d}(\rho).
\end{equation}
The quadratic Hamiltonian has the general form
\begin{align}
H  = \sum_{i,j=0}^{N-1} \msf{J}_{ij} a^\dagger_i a_j 
 + \frac{1}{2} \sum_{i,j=0}^{N-1}\msf{K}_{ij} a^\dagger_i a^\dagger_j + 
  \frac{1}{2} \sum_{i,j=0}^{N-1} \msf{K}^*_{ij} a_i a_j ,
\end{align}
where $\msf{J}$ and $\msf{K}$ are Hermitian and symmetric matrices, respectively. The Liouvillian has both loss and gain terms
\begin{eqnarray}
{\cal L}_{\rm d}(\rho) &=& 
\sum_{i,j=0}^{N-1}  \frac{\Gamma_{ij}}{2} 
\left( 2 a_i \rho a_j^\dagger - a_j^\dagger a_i \rho - \rho a_j^\dagger a_i \right) \nonumber \\
 &+& \sum_{i,j=0}^{N-1}  \frac{P_{ij}}{2} 
\left( 2 a_i^\dagger \rho a_j - a_j a_i^\dagger \rho - \rho a_j a_i^\dagger \right).
\end{eqnarray}
where both $\Gamma_{ij}$ and $P_{ij}$ are real valued Hermitian matrices. Our formalism can be straightforwardly extended to the more general case of complex $\Gamma$ and $P$, and it remains valid in the presence of long-range couplings. 

We are interested in topological properties of the steady state, $\rho^\star$, defined by
\begin{equation}
{\cal L}(\rho^\star) = 0.
\end{equation}
In previous works \cite{gomez2023driven}, the topological properties of parametric chains were characterized in terms of their linear response to coherent input signals, leading to the notion of topological amplification. In contrast, here we focus on the topological characterization of the steady state in the absence of external inputs, and on how topological features can be directly inferred from correlation functions.

We study equal-time two-point correlations in the steady state, defined as
\begin{equation}
{\cal N}_{ij}  = \langle a^\dagger_i a_j \rangle , \ \ 
{\cal M}_{ij}  = \langle a^\dagger_i a^\dagger_j \rangle.
\end{equation}
Following standard condensed-matter terminology in Bogoliubov–de Gennes (BdG) systems, we refer to ${\cal N}_{ij}$ as normal correlations,
and ${\cal M}_{ij}$ as anomalous correlations.
Equivalently, we consider normal, $\langle a_i^\dagger(\omega) a_j (\omega') \rangle$, and anomalous, $\langle a_i^\dagger(\omega) a_j^\dagger (\omega') \rangle$,    correlations in frequency space, expressed in terms of Fourier transforms of bosonic operators, $$a_j(\omega) = \frac{1}{2\pi} \int_{-\infty}^{\infty} dt e^{i \omega t} a_j(t).$$
Throughout, we use the convention
$a_j^\dagger(-\omega) \equiv [a_j(-\omega)]^\dagger$, for Fourier transforms of creation operators.

\section{Green’s function framework for evaluating bosonic correlations}\label{section2}
In this section, we outline a general framework for computing both frequency-resolved and equal-time correlations using the quantum Langevin formalism. 
This approach enables a direct characterization of steady-state topological properties through Green’s functions. We build on our previous works on topological input-output theory \cite{gomez2023driven,ramos2021topological},
which we formally recast in a convenient way to unveil the spatial structure of generic bosonic correlations.

\subsection{Quantum Langevin equation and Green's function formalism}
Bosonic operators follow the quantum Langevin equation \cite{clavero2025vibrational,gomez2023driven} given by
\begin{equation}\label{df}
\frac{d}{dt} 
\begin{pmatrix} {\boldsymbol{a}} \\ \boldsymbol{a}^\dagger \end{pmatrix} = 
-i \mathbb{H} \begin{pmatrix} \boldsymbol{a} \\ \boldsymbol{a}^\dagger \end{pmatrix} +
\begin{pmatrix} {\boldsymbol{\xi}} \\ \boldsymbol{\xi}^\dagger \end{pmatrix} ,
\end{equation}
where bold symbols denote vectors of $N$ expectation values of bosonic operators, $\mathbb{H}$ is the non-Hermitian dynamical matrix 
\begin{equation}\label{hnh}
    \mathbb{H} = 
    \begin{pmatrix} 
    \msf{J} + \frac{i}{2}\left(P-\Gamma \right) &  \msf{K} 
    \\
    - \msf{K}^*  & -\msf{J}^*  + \frac{i}{2} \left(P-\Gamma \right) 
    \end{pmatrix},
\end{equation}
that presents particle-hole-symmetry (PHS) (see App. \ref{ph}), and $\xi_j$ represent the quantum noise operators, satisfying
\begin{eqnarray}
\langle \xi_i^\dagger(t) \xi_j(t') \rangle &=& P_{ij} \delta(t-t') , \nonumber \\
\langle \xi_i(t) \xi^\dagger_j(t') \rangle &=& \Gamma_{ij} \delta(t - t').
\label{eq:langevin}
\end{eqnarray}
We are assuming that the gain/loss originates from couplings to independent baths in the vacuum state. 

We obtain a closed system of equations in frequency space for the Fourier components $\boldsymbol{a}(\omega), \boldsymbol{a}^\dagger(-\omega)$, as
\begin{equation}
- i \omega \begin{pmatrix} \boldsymbol{a}(\omega) \\ \boldsymbol{a}^\dagger(-\omega) \end{pmatrix} = 
-i \mathbb{H} \begin{pmatrix} \boldsymbol{a}(\omega) \\ \boldsymbol{a}^\dagger(-\omega) \end{pmatrix} +
\begin{pmatrix} {\boldsymbol{\xi}}(\omega) \\ \boldsymbol{\xi}^\dagger(-\omega) \end{pmatrix} .
\label{eq:frequency}
\end{equation}
Note that the frequency $\omega$ represents a detuning with respect to the coherent pump frequency of the physical model. The solution of this set of equations can be expressed in terms of the Green's function,
\begin{equation}\label{eq:frequency2}
\begin{pmatrix} \boldsymbol{a}(\omega) \\ \boldsymbol{a}^\dagger(-\omega) \end{pmatrix} = 
i \mathbb{G}(\omega) \begin{pmatrix} {\boldsymbol{\xi}}(\omega) \\ \boldsymbol{\xi}^\dagger(-\omega) \end{pmatrix},
\end{equation}
where the $2N\times 2N$ Green's function matrix ${\mathbb G}(\omega)$ is defined as
\begin{equation} 
    {\mathbb G}(\omega) \equiv \left(\omega\mathds{1}_{2N}-{\mathbb H}\right)^{-1}.
    \label{eq:green}
\end{equation}
In this work, we rely on the SVD in order to dissect the Green's function. This approach is well-suited to defining collective modes in the quantum Langevin formalism and serves as the basis for a topological input-output theory. We start by defining the SVD of the non-Hermitian dynamical matrix \cite{herviou2019defining} as
\begin{equation}
\omega \mathds{1}_{2N} - \mathbb{H} = \mathbb{U} \mathbb{S} \mathbb{V}^\dagger ,
\end{equation}
where $\mathbb{U}$ and $\mathbb{V}$ are unitary matrices and $\mathbb{S}$ is a positive diagonal matrix. 

In this representation, the Green’s function takes the form
\begin{equation} \label{svd}
\mathbb{G}(\omega) = \mathbb{V} \mathbb{S}^{-1} \mathbb{U}^\dagger .
\end{equation}
Note that singular vectors and values have an implicit frequency dependence that we are not showing to simplify the notation.
We can now define collective system operators and quantum Langevin operators as
\begin{align}
\label{eq:collective}
\begin{pmatrix} 
\boldsymbol{a}(\omega) \\ \boldsymbol{a}^\dagger(-\omega) 
\end{pmatrix} 
\equiv &
\mathbb{V} 
\begin{pmatrix} {\boldsymbol{b}}(\omega) \\ \boldsymbol{b}^\dagger(-\omega) \end{pmatrix},
\nonumber \\
\begin{pmatrix} \boldsymbol{\xi}(\omega) \\ \boldsymbol{\xi}^\dagger(-\omega) \end{pmatrix} 
\equiv&
\mathbb{U} \begin{pmatrix} {\boldsymbol{\chi}}(\omega) \\ \boldsymbol{\chi}^\dagger(-\omega) \end{pmatrix} .
\end{align}
In this new basis, the quantum Langevin equations become diagonal
\begin{equation}\label{eq:frequency3}
\begin{pmatrix} \boldsymbol{b}(\omega) \\ \boldsymbol{b}^\dagger(-\omega) \end{pmatrix} 
= 
i \mathbb{S}^{-1} \begin{pmatrix} {\boldsymbol{\chi}}(\omega) \\ \boldsymbol{\chi}^\dagger(-\omega) \end{pmatrix}.
\end{equation}
Eq \eqref{eq:frequency3} shows that the SVD yields an ideal basis for describing the amplification of quantum noise in terms of amplification channels with strength given by the inverse singular values at each frequency.

In a BdG system like ours, it is convenient to decompose the unitary matrices $\mathbb{U}$ and $\mathbb{V}$ into blocks,
\begin{eqnarray}
\mathbb{U} = \begin{pmatrix} \UU \\ \bar{\UU} \end{pmatrix}  ,
\mathbb{V} = \begin{pmatrix} \VV \\ \bar{\VV} \end{pmatrix} ,
\label{eq:blocks}
\end{eqnarray}
with $\UU_{jn} = \mathbb{U}_{jn}$ and $\bar{\UU}_{jn} = \mathbb{U}_{j+N,n}$ for $j=1,\dots,N$ and $n=1,\dots,2N$, and analogously for $\mathbb{V}$. With this block decomposition, we can express the Green's function as
\begin{equation}
\mathbb{G}(\omega) = 
\begin{pmatrix} \VV \mathbb{S}^{-1} \UU^\dagger && \VV \mathbb{S}^{-1} \bar{\UU}^\dagger 
\\ \bar{\VV} \mathbb{S}^{-1} \UU^\dagger && \bar{\VV} \mathbb{S}^{-1} \bar{\UU}^\dagger \end{pmatrix} .
\end{equation}
The above equations will allow us to extend our input-output topological formalism to the study of dissipative steady states.
\subsection{Bosonic correlations in the SVD basis}
We now study correlations using the input-output SVD formalism. Let us start with bosonic correlations in frequency space, 
\begin{eqnarray}
\langle a^\dagger_i(\omega) a_j(\omega') \rangle &=& N_{ij} (\omega) \delta(\omega-\omega'),
\nonumber \\
\langle a^\dagger_i(\omega) a^\dagger_j(-\omega') \rangle &=& M_{ij} (\omega) \delta(\omega-\omega').
\end{eqnarray}
In order to use our SVD formalism, let us write these correlations in matrix form,
such that they can be expressed in terms of the Green's function using Eqs. (\ref{eq:langevin}), (\ref{eq:frequency2}), and (\ref{svd}), leading to

\begin{eqnarray}\label{eq:corrw}
\begin{pmatrix} 
\langle \boldsymbol{a}^\dagger(\omega) \boldsymbol{a}(\omega')^T \rangle    &&   
\langle \boldsymbol{a}^\dagger(\omega) \boldsymbol{a}^\dagger(-\omega')^T \rangle
\\  
\langle \boldsymbol{a}(-\omega) \boldsymbol{a}(\omega')^T \rangle      &&   
\langle \boldsymbol{a}(-\omega) \boldsymbol{a}^\dagger(-\omega')^T \rangle  
\end{pmatrix} 
&=& {\mathbb G}^*(\omega) 
\begin{pmatrix} 
P    &&   0
\\  
0      &&   \Gamma
\end{pmatrix} 
 {\mathbb G}^T(\omega) \delta(\omega - \omega')
\nonumber \\&=& \mathbb{V}^* \mathbb{\Sigma} \mathbb{V}^T \delta(\omega - \omega').
\end{eqnarray}

In the last equation, we have used the SVD decomposition in Eq. \eqref{svd} and defined the amplification matrix,
\begin{eqnarray}
\mathbb{\Sigma} &\equiv& \mathbb{S}^{-1} \mathbb{U}^T 
\begin{pmatrix} P  &&  0\\ 0  && \Gamma \end{pmatrix} 
 \mathbb{U}^* \mathbb{S}^{-1}  =
 \mathbb{S}^{-1} 
 \left( \UU^T P \UU^* +  \bar{\UU}^T \Gamma \bar{\UU}^* \right)\mathbb{S}^{-1},
\label{eq:amplif.matrix}
\end{eqnarray}
where we have used the block decomposition in Eq. \eqref{eq:blocks} in the last equality.
From Eqs. \eqref{eq:corrw}, and \eqref{eq:amplif.matrix}, we can read the following interpretation of the steady state in terms of the singular eigensystem of the dynamical matrix: \textit{right singular vectors, $\mathbb{V}$, determine the spatial distribution of correlations at each frequency $\omega$, whereas the amplification matrix, $\mathbb{\Sigma}$, describes the propagation of the quantum fluctuations, amplified by the inverse squared singular values, $\mathbb{S}^{-1}$}. Note that left singular vectors, $\mathbb{U}$, are fully contracted within the amplification matrix.
It is also clear from our result that singular vectors with small singular values dominate correlation functions.

We finally arrive to the following compact representation of the two-point frequency-resolved correlations,
\begin{eqnarray}
N(\omega) &=& 
\VV^* \mathbb{\Sigma}  \VV^T ,
\nonumber \\
M(\omega) &=& \VV^* \mathbb{\Sigma} \bar{\VV}^T .
\label{eq.NM}
\end{eqnarray}
We will make use of these expressions for numerical calculations in the following sections. 

Equal-time two-point correlation functions can then be obtained as
\begin{eqnarray}
\label{freq.integrated}
{\cal N}_{ij} &=& \frac{1}{2\pi} \int d\omega N_{ij}(\omega)  , 
\nonumber \\
{\cal M}_{ij} &=& \frac{1}{2\pi} \int d\omega M_{ij}(\omega) .
\end{eqnarray}
These integrated correlation functions can be more compactly expressed by defining the correlation matrix,
\begin{equation}\label{hnh}
    \mathcal{C} = 
    \begin{pmatrix} 
    {\cal N} &  {\cal M} 
    \\
    {\cal M}^*  & {\cal N} + \mathds{1}_N
    \end{pmatrix}.
\end{equation}
Using Eqs. (\ref{eq:corrw}), and (\ref{freq.integrated}),  we can derive the expression
\begin{equation} \label{C.convolution}
\mathcal{C} = \int \frac{d\omega}{2 \pi}  
\mathbb{G}^*(\omega)  \begin{pmatrix}
P && 0 \\
0 && \Gamma
\end{pmatrix}  \mathbb{G}^t(\omega) .
\end{equation}
Eq. \eqref{C.convolution} can be obtained independently from the master equation \cite{clavero2025vibrational} or Keldysh formalism \cite{gomez2022bridging}. 

We introduce normalized frequency-resolved and equal-time correlations as \begin{equation}\label{Nnorsm}
    \bar{{ N}}_{i j}(\omega)=\frac{{ N}_{ij}}{\sqrt{{ N}_{ii}{\ N}_{jj}}}, \qquad \bar{{\cal N}}_{i j}=\frac{{\cal N}_{ij}}{\sqrt{{\cal N}_{ii}{\cal N}_{jj}}}.
\end{equation}
Equivalently, the normalized anomalous correlations are defined as
\begin{equation}\label{Mnorm}
    \bar{{ M}}_{i j}(\omega)=\frac{{ M}_{ij}}{\sqrt{{ N}_{ii}{\ N}_{jj}}}, \qquad \Bar{{\cal M}}_{i j}=\frac{{\cal M}_{ij}}{\sqrt{{\cal N}_{ii}{\cal N}_{jj}}}.
\end{equation}
The normalization in Eqs. (\ref{Nnorsm}, \ref{Mnorm}) ensures that correlations are evaluated relative to the local excitation number at each site. These normalized correlators will be used below to characterize long-range order, defined as 
\begin{equation}\label{LROdef}
    \lim_{|i-j|\rightarrow \infty}\bar{N}_{i j}(\omega)\neq 0,
\end{equation}
 which constitutes a convenient observable to assess the spatial extent of correlations. 

\section{Physical bosonic dissipative models}\label{section3}
We introduce two driven-dissipative quantum chains that serve as the underlying models 
from which correlation properties within topological phases are extracted. 

\subsection{Model I: Dissipative BdG chain}\label{modelI}  
Our first model is a bosonic dissipative BdG chain, with the following Hermitian matrices
\begin{align}
       \msf{J}_{ij} &= \Delta\delta_{ij}+ J (e^{i \phi}\delta_{i,j+1}+e^{-i \phi}\delta_{i,j-1}) , \nonumber \\
       \msf{K}_{ij} &= g_s\delta_{ij}+g_c(\delta_{i,j+1}+\delta_{i,j-1}) , \nonumber \\
       \Gamma_{ij}  &= \gamma \delta_{ij} .
       \label{eq:g}
\end{align}
We include boson tunneling terms, $J$, single-site parametric couplings, $g_s$, and nearest-neighbour parametric couplings, $g_c$, together with local photon decay $\gamma$.
This model is a variation of the Bosonic Kitaev chain introduced in \cite{mcdonald2018phase} and studied in the context of topological amplification in  \cite{gomez2023driven}. It can be implemented with superconducting circuit arrays that break time-reversal symmetry \cite{Ramos:2022kvf}, and with trapped ions in individual microtraps \cite{clavero2025vibrational}.
\begin{figure}[h]
    \centering
    \includegraphics[width=0.75\textwidth]{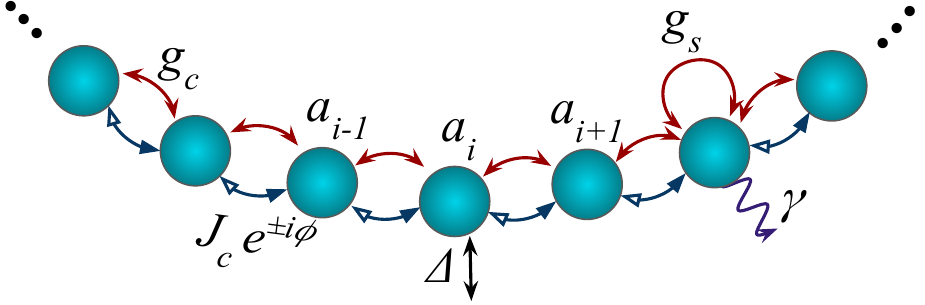}
    \caption{Schematic representation of the dissipative bosonic Kitaev chain. Each site has onsite energy $\Delta$, single-site parametric term $g_s$, and local losses $\gamma$. Neighboring sites are connected by non-reciprocal complex hopping $Je^{\pm i\phi}$ (blue arrows),  and parametric couplings $g_c$ (red arrows). }
    \label{schemes}
\end{figure}

Without loss of generality, and throughout this work, we focus on Model I in a highly symmetric parameter regime defined by 
$J = g_s = g_c = g$, where $g$ is the common energy scale, $\Delta=0$, and $\phi=\pi/2$, with $\gamma$ as the only tunable parameter left.

\subsection{Model II:  Dimerized dissipative BdG chain}\label{modelII}

\begin{figure}[h]
    \centering
    \includegraphics[width=0.75\textwidth]{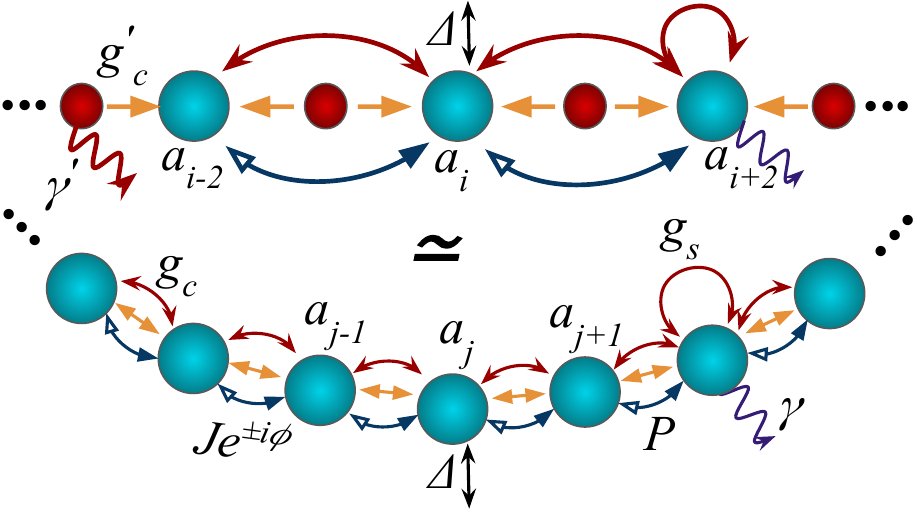}
    \caption{Top: Dimerized version of the dissipative bosonic Kitaev chain. Even (blue) sites correspond to the main chain, characterized by onsite energy $\Delta$, non-reciprocal complex hopping $Je^{\pm i\phi}$ (blue arrows), local ($g_s$) and nearest-neighbor ($g_c$) parametric couplings (red arrows), and local losses $\gamma$. Odd (red) sites represent auxiliary modes, coupled to the main chain via $g_c'$, and subject to much stronger dissipation $\gamma' \gg \gamma$. Bottom: Upon adiabatic elimination of the auxiliary sites, the system maps onto an effective bosonic chain equivalent to Model I, featuring collective gain terms  $P_{jl}$ (yellow arrows).}
    \label{schesmes}
\end{figure}

Our second model is motivated by the search for more complex topological phases that will follow in the upcoming sections. 
In previous work \cite{gomez2023driven}, we found that non-local incoherent pumping can lead to winding numbers greater than one. Here, we present a dimerized version of the BdG model in which collective gain terms appear from the adiabatic elimination of a set of auxiliary sites,
\begin{align}
       \msf{J}_{ij} &= \Delta\delta_{ij \in {\rm even}}+ J (e^{i \phi}\delta_{i,j+2}+e^{-i \phi}\delta_{i,j-2})_{{ij} \in {\rm even}} , \nonumber \\
       \msf{K}_{ij} &= g_s\delta_{ij \in {\rm even}} + 
       g_c \left( \delta_{i,j+2} +\delta_{i,j-2} \right)_{{ij} \in {\rm even}} + g_c'(\delta_{i,j+1}+\delta_{i,j-1}) , 
       \nonumber \\
       \Gamma_{ij}  &= \gamma \delta_{ij \in {\rm even}} + \gamma' \delta_{ij \in {\rm odd}} .
       \label{eq:g}
\end{align}
In this model, even sites are system sites with decay $\gamma$, whereas odd sites are auxiliary sites with a different decay rate, $\gamma'$.
Even sites are coupled via complex photon-tunneling and parametric terms with amplitudes $J$ and $g_c$, respectively. Even and odd sites are coupled by nearest-neighbor parametric couplings, $g_c'$. Our goal is to obtain, after this adiabatic elimination, a driven-dissipative effective model with collective gain terms (see App. \hyperref[app:anexoA]{B}). 
Model II is depicted in Fig. \ref{schesmes} (top), where even (main) modes are presented in blue, while the odd (auxiliary) modes are depicted in red. 

The resulting coupling with the auxiliary sites gives rise to an extra effective Lindbladian term for even sites,
\begin{equation}
    \mathcal{L}_p({\rho})=\sum_{j,l \in {\rm even}}^{N}\msf{P}_{jl}(2a^\dagger_j\rho_a a_l -a_ja^\dagger_l \rho_a - \rho_a  a_ja^\dagger_l),
\end{equation}
with the collective gain matrix represented as
\begin{equation}
\msf{P}_{jl} =\frac{2 g_c'^2}{\gamma'}(2\delta_{jl} + \delta_{j,l+2} + \delta_{j,l-2}) .
\end{equation}
Note that the effective model is represented as a chain with collective gain acting on the even sites. Since the odd sites are adiabatically eliminated, each even site of the original system is mapped onto a single site labeled by a natural number in the effective model, as illustrated in Fig.~\ref{schesmes} (bottom).

This approach provides an experimentally feasible pathway to achieve the nonlocal pumping required to access topological phases with large winding numbers \cite{gomez2023driven}. Throughout this work, we focus on Model II in the parameter regime defined by $J=g$, $g_s=g_c=0.1g$,  $\Delta=0$, and $\phi=\pi/2$, with $\gamma$, $g_c'$, and $\gamma'$ as tunable parameters left.

\section{Topological amplification and invariants}\label{section4}
In this section, we review the theory of topological amplification and employ it to classify the topology of steady states in bosonic quadratic dissipative chains.

\subsection{Spectral non-Hermitian winding number and connection to SVD}

The SVD is the starting point for a topological characterization of steady states of quadratic Liouvillians. 
Firstly, through the hermitization method, \cite{feinberg1997non}, we embed the non-Hermitian dynamical matrix, $\mathbb{H}$, 
into an extended Hermitian matrix, doubling the Hilbert space dimension, 
\begin{eqnarray}
   \mathcal{H}(\omega) &\equiv &\begin{pmatrix}0 & \omega \mathbb{1} - \mathbb{H} \\\omega \mathbb{1} - \mathbb{H}^\dagger&0\end{pmatrix} .
\end{eqnarray}
A crucial observation is that the eigensystem of the extended Hermitian matrix ${\cal H}(\omega)$ is the SVD of the non-Hermitian dynamical matrix
$\omega \mathbb{1}-\mathbb{H}$ \cite{porras2019topological},
\begin{equation}\label{HUS}  
\mathcal{H}(\omega)\begin{pmatrix}\mathbb{U}\\ \mathbb{V}\end{pmatrix} 
= 
\begin{pmatrix} \mathbb{U} \mathbb{S}\\ \mathbb{V}\mathbb{S} \end{pmatrix}.
\end{equation}
We notice that ${\cal H}(\omega)$ has an intrinsic chiral symmetry 
\begin{equation}
\sigma_z {\cal H}(\omega) \sigma_z = - {\cal H}(\omega) .
\label{eq:chiral}
\end{equation}
The property established by Eq. \eqref{eq:chiral} implies that zero energy states (and, thus, zero singular vectors of $\omega \mathbb{1} - \mathbb{H}$) are topologically protected. From now on, we name such states {\it topological zero singular vectors}. 
Zero singular vectors are the dominant contribution to the Green's function, since the corresponding zero singular values appear inverted, as shown in  Refs.~\cite{porras2019topological,gomez2023driven}.  
The number of existing topological zero singular vectors and values is determined by the spectral winding number (hereafter referred to simply as winding number)
\begin{equation}\label{winding1}
\nu(\omega)
= \Im \int_{-\pi}^{\pi}
\frac{dk}{2\pi}
\Tr \partial_k \log \left( \omega \mathds{1}_{2N} - \mathbb{H}(k) \right),
\end{equation}
where $\mathbb{H}(k)$ denotes the non-Hermitian dynamical matrix under periodic boundary conditions (PBC), expressed in the plane-wave basis. 

\begin{figure}[t]
    \centering
    \includegraphics[width=0.75\textwidth]{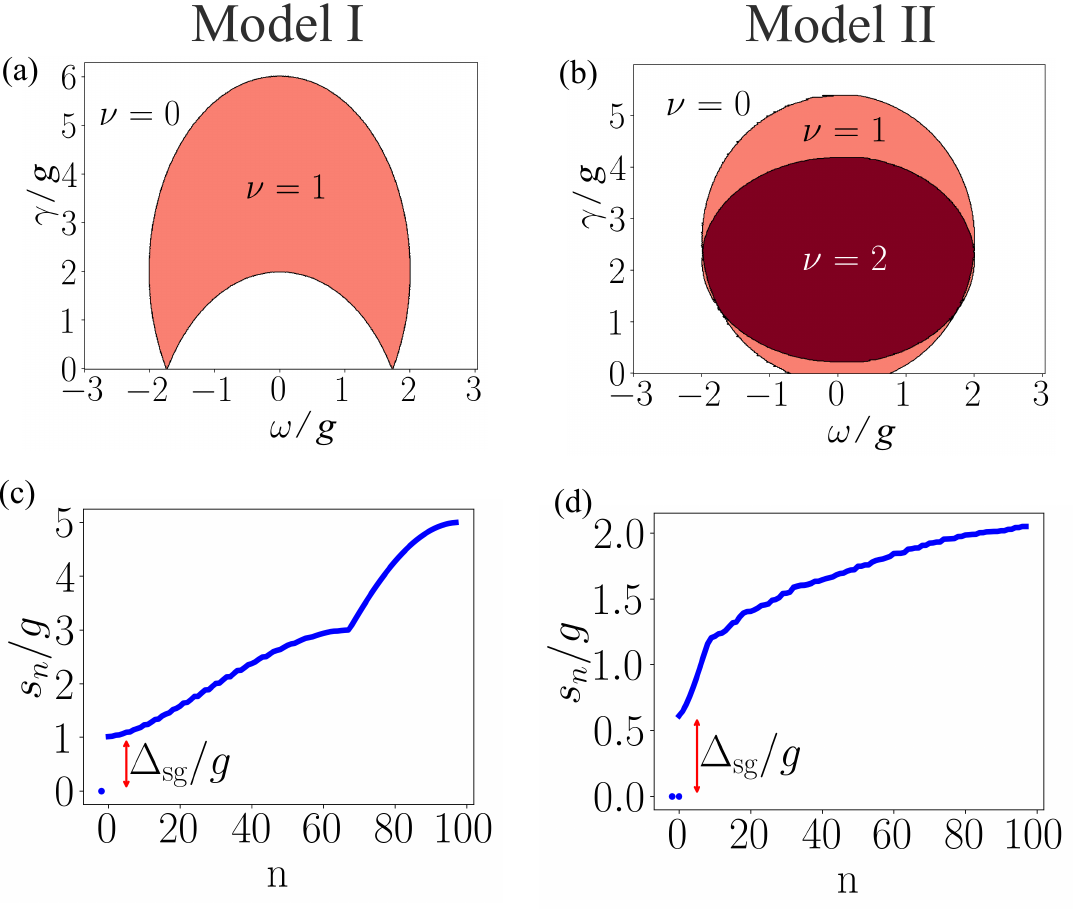}
    \caption{(a) Model I: Topological phase diagram for the symmetric parameter set defined in Sec.~\ref{modelI}. (b) Model II: Topological phase diagram for  with $g_s/g=g_c/g=0.1$, $g_c'/g=3$ and $\gamma'/g=30$ (corresponding to $P/g=\frac{3}{5}$). (c) Singular value spectrum under OBC in a nontrivial region of (a) with $\nu=1$, for $\gamma/g=4$ and $\omega/g=0$ in (a), exhibiting a unique zero singular value separated from the bulk by a gap $\Delta_{\rm sg}$.   (d) Singular value spectrum in a nontrivial region of (b) with $\nu=2$, for $\gamma/g=3$, and $\omega/g=0$, exhibiting a couple of zero singular values. The system size is $N=50$.}
    \label{cuadro}
\end{figure}

Non-Hermitian topological phases, identified by a nonzero winding number $\nu(\omega) \neq 0$, signal the emergence of zero singular values and exponentially localized edge singular vectors under open boundary conditions (OBC). For instance, Fig.~\ref{cuadro} shows the topological phase diagrams for Models I and II, where a phase with winding number equal to one (two) gives rise to one (two) zero singular values, separated from the bulk by a gap  $\Delta_{\rm sg}$. These modes can be regarded as the dissipative analogues of zero-energy boundary states in Hermitian topological insulators \cite{hasan2010colloquium}. In the topological regime, the corresponding singular values vanish exponentially in the thermodynamic limit, thereby dominating the behavior of the Green’s function in Eq.~\eqref{svd}. This allows us to approximate
\begin{equation}
\mathbb{G}_{\mu \nu}(\omega) \approx \sum_{n \in \mathrm{E}} \mathbb{V}_{\mu n} \mathbb{S}^{-1}_n \mathbb{U}_{\nu n}^* ,
\end{equation}
where the sum runs over the subset $\mathrm{E}$ of topological zero singular values.

\subsection{Topological invariant for steady-states}
\begin{figure}[t]
    \centering
    \includegraphics[width=0.65\textwidth]{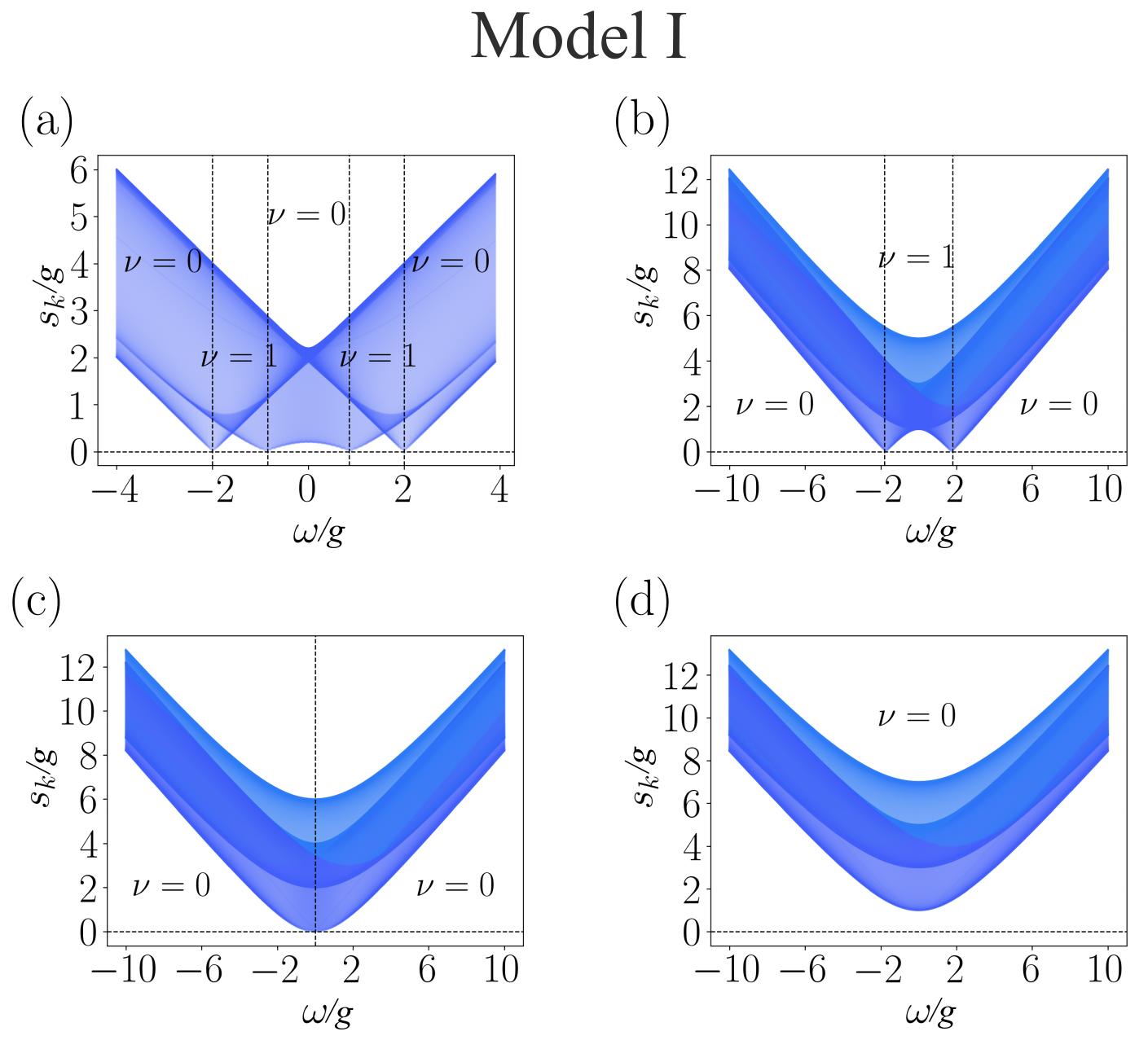}
    \caption{Model I: Singular values as a function of the frequency $\omega$ for PBC. 
    (a) Topologically non-trivial phase with $\Vec{\mathcal{V}_a}=(0,1,0,1,0)$, for $\gamma/g=1.6$.
    (b) Topologically non-trivial phase with $\Vec{\mathcal{V}_b}=(0,1,0)$, for $\gamma/g=4$. 
    (c) Critical point with $\Vec{\mathcal{V}_c}=(0,0)$, for $\gamma_c/g=6$. 
    (d) Trivial phase with $\Vec{\mathcal{V}_d}=(0)$, for $\gamma/g=8$. Each component of $\Vec{\mathcal{V}}$ indicate the topological winding numbers associated with distinct frequency intervals. For example, $\Vec{\mathcal{V}_a}=(0,1,0,1,0)$ corresponds to alternating regions of trivial and non-trivial topology, separated by singular value gap closings, while $\Vec{\mathcal{V}_b}=(0,1,0)$ indicates a single non-trivial frequency band flanked by trivial regions. 
    The parameters correspond to the symmetric point defined in Sec.~\ref{modelI}. }
    \label{PBC}
\end{figure}

\begin{figure}[h]
    \centering
    \includegraphics[width=0.65\textwidth]{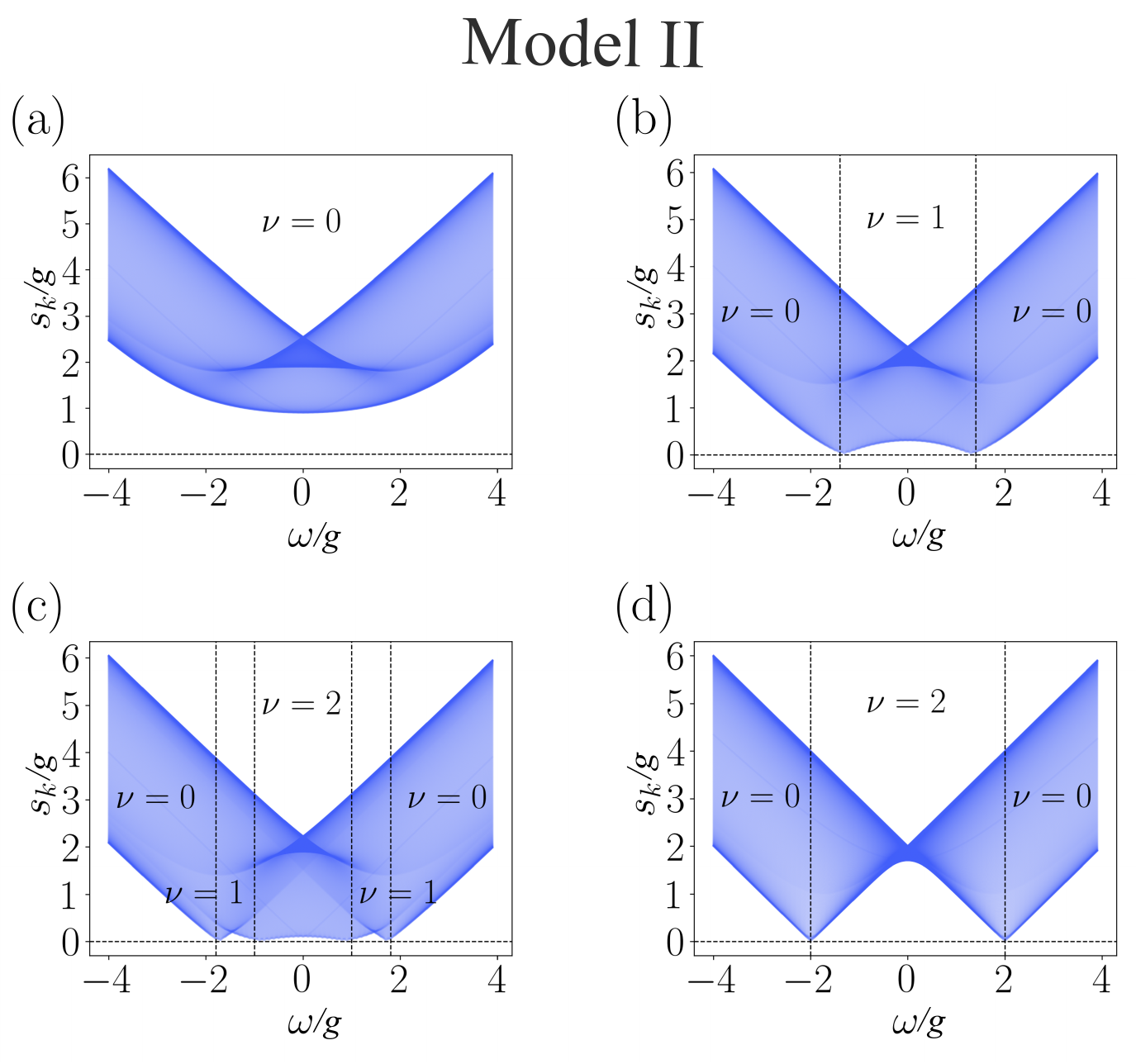}
    \caption{Model II: Singular values as a function of the frequency $\omega$ for PBC. 
    (a) Trivial phase with $\Vec{\mathcal{V}_a}=(0)$, for $g_c'/g=2$, $\gamma'/g=20$ (corresponfing to $P/g=\frac{2}{5}$).
    (b) Topologically non-trivial phase with  $\Vec{\mathcal{V}_b}=(0,1,0)$, for $g_c'/g=2.5$, $\gamma'/g=25$ ($P/g=0.5$). 
    (c) Topologically non-trivial phase with $\Vec{\mathcal{V}_c}=(0,1,2,1,0)$, for $g_c'/g=3$, $\gamma'/g=30$ ($P/g=\frac{3}{5}$). 
    (d) Topologically non-trivial phase with  $\Vec{\mathcal{V}_d}=(0,2,0)$, for $g_c'/g=5$, $\gamma'/g=50$ ($P/g=1$). 
    The remaining parameters are defined in Sec. \ref{modelII}, and $\gamma/g=4$.}
    \label{PBC2}
\end{figure}

The winding number $\nu(\omega)$ in Eq. \eqref{winding1} is a natural starting point to characterize topological steady state properties of bosonic quadratic Liouvillians. 
However, the topological phase of steady states in the absence of external monochromatic driving is not uniquely characterized at a single frequency. To illustrate this point, Figs.~\ref{PBC} and \ref{PBC2} show the singular value spectrum under PBC as a function of $\omega$ for different parameter regimes, together with the corresponding topological regions, for Models I and II, respectively. 
In these plots, we find regions in frequency space with different winding numbers, separated by distinct closures of singular-value gaps.

Since there is no any global unique value of the winding number, but it is a function of frequency, the question arises of 
{\it how to extend the concepts of topological equivalence and topological invariants to quadratic Liouvillians.} 
For this, we briefly review the concept of adiabatic deformation of Hamiltonians in topological insulator theory
\cite{asboth2016short}.
Assume we have a translationally invariant Hamiltonian $H_\lambda$ that depends on a control parameter $\lambda$. 
Two gapped Hamiltonians $H_{\lambda_1}$ and $H_{\lambda_2}$ are topologically equivalent if they can be adiabatically deformed into each other, that is, $H_{\lambda_1}$ can be transformed into $H_{\lambda_2}$ by continuously varying $\lambda$ without closing the energy gap. 

In the case of quadratic Liouvillians, the concept of adiabatic deformation between two parametrized Liouvillians, ${\cal L}_{\lambda_1}$ and ${\cal L}_{\lambda_2}$, or equivalently, parametrized non-Hermitian dynamical matrices, 
$\mathbb{H}_{\lambda_1}$ and $\mathbb{H}_{\lambda_2}$
is more subtle. 
The singular value gap is not a single value but a function $\Delta_{sg}(\omega)$ that vanishes at certain points. A small perturbation will only affect those values such that 
$\Delta_{sg}(\omega) \approx 0$. 
To prove this, consider a small perturbation 
$\mathbb{H}_\lambda \to \mathbb{H}_{\lambda + \delta \lambda}$. Weyl's inequality \cite{horn2012matrix}
applied to singular values can be used to bound the perturbation of the singular value gap, 
\begin{equation}
|\Delta'_{sg}(\omega) - \Delta_{sg}(\omega) |\leq \| \mathbb{H}_{\lambda + \delta \lambda} - \mathbb{H}_\lambda \|_2,
\end{equation}
where $\Delta_{sg}'(\omega)$ and $\Delta_{sg}(\omega)$ are the singular gaps of 
$\omega  \mathds{1}_{2N} - \mathbb{H}_{\lambda+\delta\lambda}$ and 
$\omega  \mathds{1}_{2N} - \mathbb{H}_\lambda$, respectively, and 
$\| A \|_2$ is the spectral 2-norm of $A$. Thus, as we perturb the quadratic Liouvillian, only those frequencies close to gap closures 
$\Delta_{sg}(\omega_j) = 0$ will be affected.

\begin{figure}[t]
    \centering
    \includegraphics[width=0.75\textwidth]{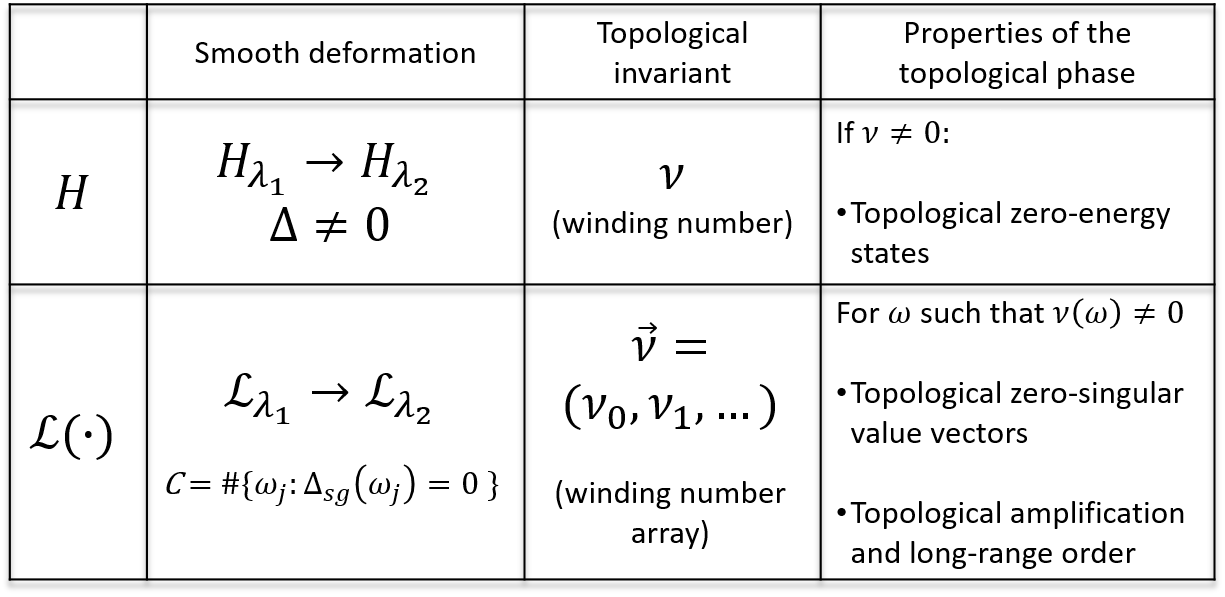}
    \caption{Comparison of topological equivalence under smooth deformations in closed-system Hamiltonians and quadratic Liouvillians of open quantum systems. In closed systems, smooth deformations of topologically protected Hamiltonians preserve the spectral gap and are characterized by a nontrivial winding number, ensuring the existence of zero-energy topological states. In quadratic Liouvillians, smooth deformations preserve the number of singular-value gap closings along the frequency axis and are described by a winding-number array, whose components encode the winding between successive gap-closing points. This invariant guarantees the presence of zero singular values and corresponding singular vectors, and underlies observable signatures such as topological amplification and long-range order.}
    \label{summ_winding}
\end{figure}

Following this observation, {\it we define the adiabatic deformation of a quadratic Liouvillians, as a continuous transformation from 
${\cal L}_{\lambda_1}$ to ${\cal L}_{\lambda_2}$, or, equivalently, from $\mathbb{H}_{\lambda_1}$ to $\mathbb{H}_{\lambda_2}$, 
such that the number of frequencies $\omega_j$ at which the singular gap closes remains constant}. 
With this definition, two quadratic Liouvillians are topologically equivalent if they can be adiabatically transformed into each other. A visual comparison between topology in closed-system Hamiltonians and quadratic Liouvillians is presented in Fig.~\ref{summ_winding}.

Our notion of adiabatic deformation leads to a definition of a topological invariant associated to topologically equivalent classes of quadratic Liouvillians in terms of a winding number array, $\Vec{\mathcal{V}}$.
The components of $\Vec{\mathcal{V}}$ encode the spectral winding number, $\nu(\omega)$, in each frequency interval, $\Omega_i\subset \mathbb{R}_\omega$, separated by singular value gap closings, i.e., 

\begin{eqnarray}
\Vec{\mathcal{V}} \equiv \big( \nu(\Omega_0), \nu(\Omega_1), \dots, \nu(\Omega_i), \dots, \nu(\Omega_n) \big). 
\end{eqnarray}
Adiabatically deforming a quadratic Liouvillian cannot change $\Vec{\mathcal{V}}$ since this would involve the emergence/disappearance of zeros of $\Delta_{sg}(\omega)$.

The winding number array $\Vec{\mathcal{V}}$ shows the following features:
\begin{itemize}
    \item The limit for large frequency values $\omega \rightarrow\pm \infty$ leads to a null winding number $\nu(\omega) = 0$. This can be directly observed from Eq. (\ref{winding1}) where the identity contribution dominates.
    \item $\mathbb{H}$ presents a particle-hole symmetry which implies that the singular value spectrum of $\omega \mathbb{1}- \mathbb{H}$ is an even function in 
    $\omega$ (i.e., $s_i(\omega)=s_i(-\omega)$). This is shown in App. \ref{phsdm}, and leads to the following symmetric structure of the winding number array: \begin{equation}
        \Vec{\mathcal{V}}=\left(\nu(-\Omega_0),\nu(-\Omega_1),...,\nu(\Omega_{k}),...,\nu(\Omega_1),\nu(\Omega_0) \right),
    \end{equation} where $\nu(-\Omega_i)=\nu(\Omega_i)$. Besides, $\nu(\pm\Omega_0)=0$ by the previous argument.
    \item  A topological phase transition is defined as a change in the dimension of the winding number array $\Vec{\mathcal{V}}$. Winding number arrays must be odd-dimensional due to the symmetry in $\omega$ in order to be in a stable topological phase. 
    Even-dimensional winding vectors correspond to critical points where two gap-closing points coalesce (see, for example, Fig. \ref{PBC} (c)). Such critical points are unstable against perturbations, since a perturbation leads to the appearance of two zeros of $\Delta_{sg}(\omega)$ (Fig. \ref{PBC} (b)), or the opening of the gap, Fig. \ref{PBC}(d).
\end{itemize}

We then assert that two integrated systems, given by $\Vec{\mathcal{V}}_1$ and $\Vec{\mathcal{V}}_2$, are topologically equivalent if and only if: (i) both vectors presents the same dimension, and (ii) $\nu_1(\Omega_i)=\nu_2(\Omega_i' )$ for every $i$. Taking into account this definition, we observe that the four illustrative examples in Fig. \ref{PBC} are topologically inequivalent since $\Vec{\mathcal{V}}_{a}=(0,1,0,1,0)$, $\Vec{\mathcal{V}}_{b}=(0,1,0)$, $\Vec{\mathcal{V}}_{c}=(0,0)$, and $\Vec{\mathcal{V}}_{d}=(0)$ have different dimensions. The even dimension and redundancy of (c) indicates that the system undergoes a phase transition. Something similar occurs in Fig. \ref{PBC2}, where we can explore more exotic non-trivial phases with winding numbers larger than one.

\section{Topologically protected correlations} \label{section5}
In this section, we demonstrate the enhancement of both frequency-resolved and equal-time two-point correlations in nontrivial topological phases, consistent with previous numerical observations \cite{rassaert2025emerging}. In this sense, singular-value topology is not merely a formal construction, but acquires a direct operational meaning in terms of measurable correlation functions.

\subsection{Topological two-point long-range correlations in frequency space}\label{6.1}
The existence of topological zero singular vectors allows us to simplify the expression for two-point correlators in topologically nontrivial regimes. 
Eqs. (\ref{eq:amplif.matrix},  \ref{eq.NM}) imply that the amplification matrix
$\mathbb{\Sigma}$ will be dominated by inverse values of zero topological singular values. Assuming an homogeneous, finite size chain with $N$ sites, such singular values contribute in \eqref{eq:amplif.matrix} with factors 
\begin{equation}
\frac{1}{\mathbb{S}_n} \propto e^{N/\xi_n}, \ \  n \in \rm E,
\end{equation}
where $\rm E$ is the subset of topological zero modes, and $\xi_n$ are the localization lengths of topological edge-states. Two-point frequency-resolved correlations can thus be approximated as
\begin{align}\label{approxNM}
N_{lj}(\omega) 
& \approx \sum_{n \in E, m \in E} \VV^*_{l n} 
\mathbb{\Sigma}_{nm} \VV_{jm}, \nonumber \\
M_{lj}(\omega) 
& \approx \sum_{n \in E, m \in E} 
\VV^*_{l n} \mathbb{\Sigma}_{nm} \bar{\VV}_{jm},
\end{align}
that is, zero-singular values lead to large values in $\Sigma_{nm}$ that dominate the sum. This result is general for driven-dissipative BdG systems with independence of the particular model.

In the case where there is a single (labeled by $n,m = 0$) topological zero singular value state, or only one of them dominates the summation, the situation further simplifies and corelation functions become, simply, 
\begin{eqnarray} \label{zerosing}
N_{jl}(\omega) &\approx& \VV^*_{l0} \mathbb{\Sigma}_{00} \VV_{j0},
\nonumber \\
M_{jl}(\omega) &\approx& \VV^*_{l0} \mathbb{\Sigma}_{00} \bar{\VV}_{j0}.  
\end{eqnarray}
The above equation implies that dominance of a single topological singular vector leads to a condensation effect into a single bosonic mode, leading to the emergence of off-diagonal long-range order \cite{dalfovo1999theory}.

This effect is even more apparent in the normalized correlations, in agreement with Eq. (\ref{LROdef}). Defining the topological edge modes in terms of amplitude and phase like $V_{j0} \equiv |V_{j0}| e^{i \phi_0(j)}$, 
$\bar{V}_{j0} \equiv |\bar{V}_{j0}| e^{i \bar{\phi}_0(j)}$
leads to the following normalized correlation functions
\begin{eqnarray}\label{eq:n1}
\bar{N}_{lj}(\omega) \approx e^{-i(\phi(l) - \phi(j))}.
\end{eqnarray}
Equivalently, for the anomalous normalized correlation, we obtain
\begin{eqnarray}\label{eq:m1}
\bar{M}_{lj}(\omega) \approx 
\frac{|\bar{\VV}_{l0}|}{|\VV_{j0}|} 
e^{-i(\phi(l) - \bar{\phi}(j))}. 
\end{eqnarray}
This equation can be further simplified using symmetries, for example, in Model I, under conditions $\phi = \pi/2$, $\Delta = 0$, we can show that $|\VV_{l0}| = |\bar{\VV}_{l0}|$ (see Appendix \ref{SPR}), such that we have full-synchronization in anomalous correlations.

Eqs. (\ref{eq:n1}, \ref{eq:m1}) are one of the key results of our paper. {\it Topological amplification phases yield long-range topological correlations in frequency space. Our results can also be understood in terms of quantum synchronization between distant sites}. Normal correlation functions $\bar{N}_{jl}$ acquire a long-range order structure, whereas $\bar{M}_{jl}$ get an additional weight related to the BdG structure of the Hamiltonian. The latter can be simplified, for example in the special symmetric case. 

\begin{figure}[h]
    \centering
    \includegraphics[width = 0.75\textwidth]{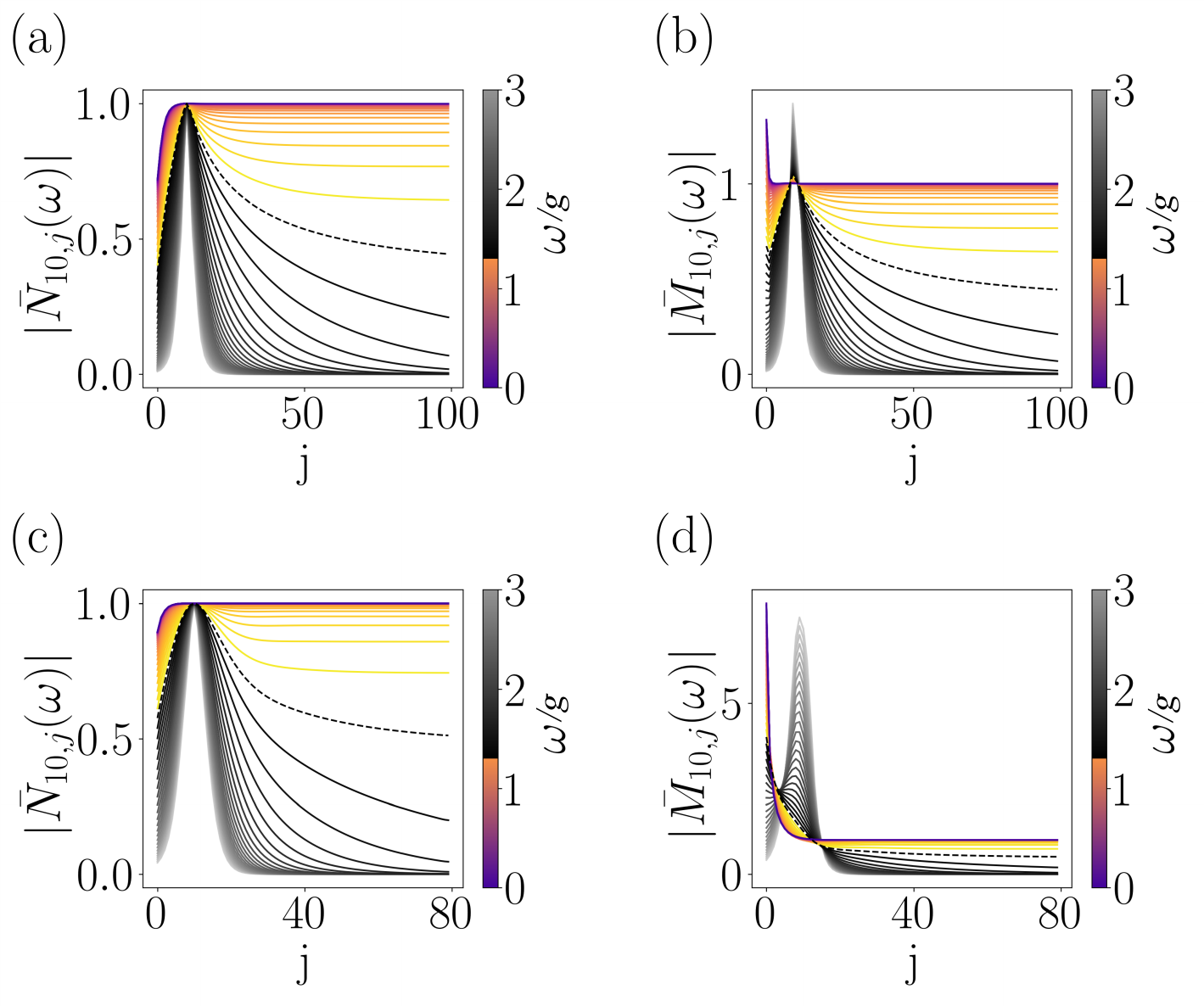}
    \caption{Normalized frequency-resolved correlations as a function of the site index $j$, for a fixed reference site $i=10$.  Colored/black curves indicate topological/trivial phases. 
    The black dashed line marks the critical point separating both regimes. (a, b) Model I, in the symmetric parameter regime defined in Sec. \ref{modelI}, with $N=100$, and $\gamma/g = 5$. 
    (c, d) Model II in the parameter regime defined in Sec. \ref{modelII}, with $N=80$, $\gamma/g=3.75$, $g_c'/g=3$, and $\gamma'/g=30$ (corresponding to $P/g=\frac{3}{5}$).}
    \label{nmb1}
\end{figure}

This interpretation holds when the system is in a topological phase with winding number $\nu(\omega) = \pm1$, where just a single singular value is strongly suppressed compared to the rest. In contrast, for topological regimes with higher winding numbers, several singular-value channels compete, and the previous expressions break down when the zero singular values are of comparable magnitude \cite{brunelli2023restoration}. Nonetheless, our situation in Model II is easier to analyze since: two zero singular values are separated from the bulk by a finite gap, while one of them remains significantly more suppressed than the others, as we can observe in Fig. \ref{s1s2} of App. \ref{sv}. In this case, previous Eqs. (\ref{eq:n1}, \ref{eq:m1}) remain valid, and, regarding long-range order, the system becomes indistinguishable across topological phases of different non-trivial winding numbers.

The normalized correlations in Fig. \ref{nmb1}  remain finite at large distances within the topological phase (colored curves), signaling the emergence of long-range order. By contrast, in the trivial phase (black curves), the normalized correlations are short-ranged and decay exponentially as a function of distance.

\begin{figure}[h]
    \centering
    \includegraphics[width = 0.75\textwidth]{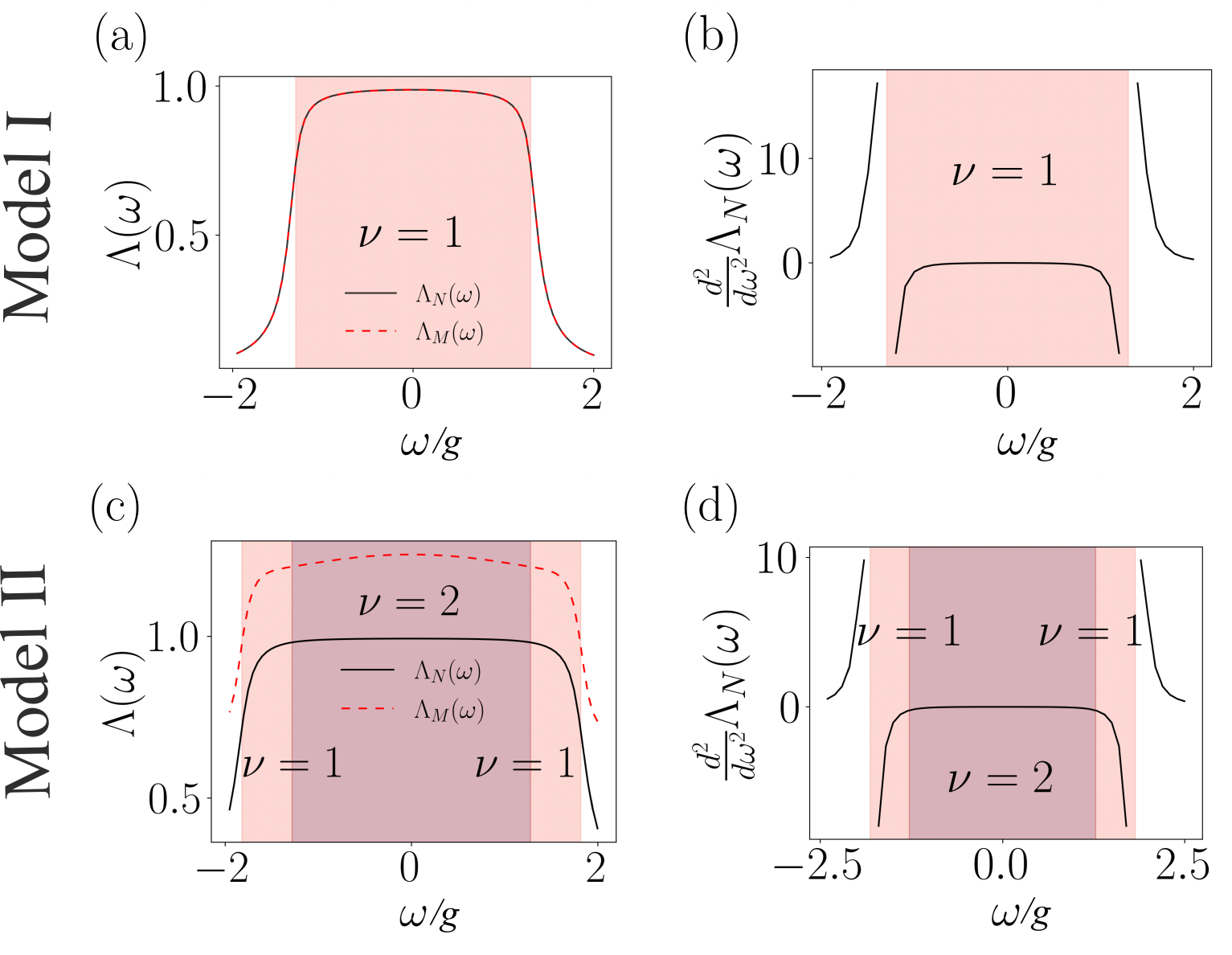}
    \caption{LRO  parameter as a function of the frequency $\omega$, for normal ($\Lambda_N$, solid black) and anomalous ($\Lambda_M$, dashed red) frequency-resolved correlations. In Model I (panel a), the LRO shows a one-to-one correspondence with the topological phase $\omega/g \in [-1.3,1.3]$. A similar behavior is observed in Model II (panel c), although the topological phase with $\nu=2$ ($\omega/g \in [-1.28,1.28]$) is indistinguishable from $\nu=1$  ($\omega/g \in [-1.82,1.28]\cup [1.28,1.82] $) based on the correlations. Panels (b) and (d) display the second derivative of the global LRO parameter, highlighting divergences at the critical points. All parameters are as in Fig. \ref{nmb1}.}
    \label{cuadrold}
\end{figure}

To quantify the presence of long-range order across the phase diagram, we introduce the LRO parameter
\begin{equation}
    \Lambda_N(\omega)=\sum_{ij}\frac{|\bar{N}_{ij}(\omega)|}{N^2},
\end{equation}
which provides a single measure of non-decaying correlations at a given frequency. We can obtain a similar quantity for the normalized anomalous frequency-resolved correlations $\bar{M}(\omega)$. Mapping $\Lambda(\omega)$ as a function $\omega$ reveals a direct correspondence between topologically non-trivial phases and finite long-range order (see Fig. \ref{cuadrold}). Note that at the critical point where the system undergoes a transition from the trivial to the topological phase, the second derivative of the LRO parameter diverges, indicating a second-order phase transition. However, we do not observe any transition from $\nu=1$ to $\nu=2$ in \ref{cuadrold} (c).

\subsection{Robustness to disorder and LRO}
A defining feature of a topologically nontrivial phase is its robustness against disorder \cite{hasan2010colloquium,bansil2016colloquium}, such that the properties of the clean system persist up to a critical disorder strength $W_c$. 
    As discussed in the previous section, one manifestation of topology in the steady state is the emergence of LRO in frequency-resolved two-point correlations. In this section, we investigate the robustness of this effect to disorder.

\begin{figure}[h]
    \centering
    \includegraphics[width=0.7\textwidth]{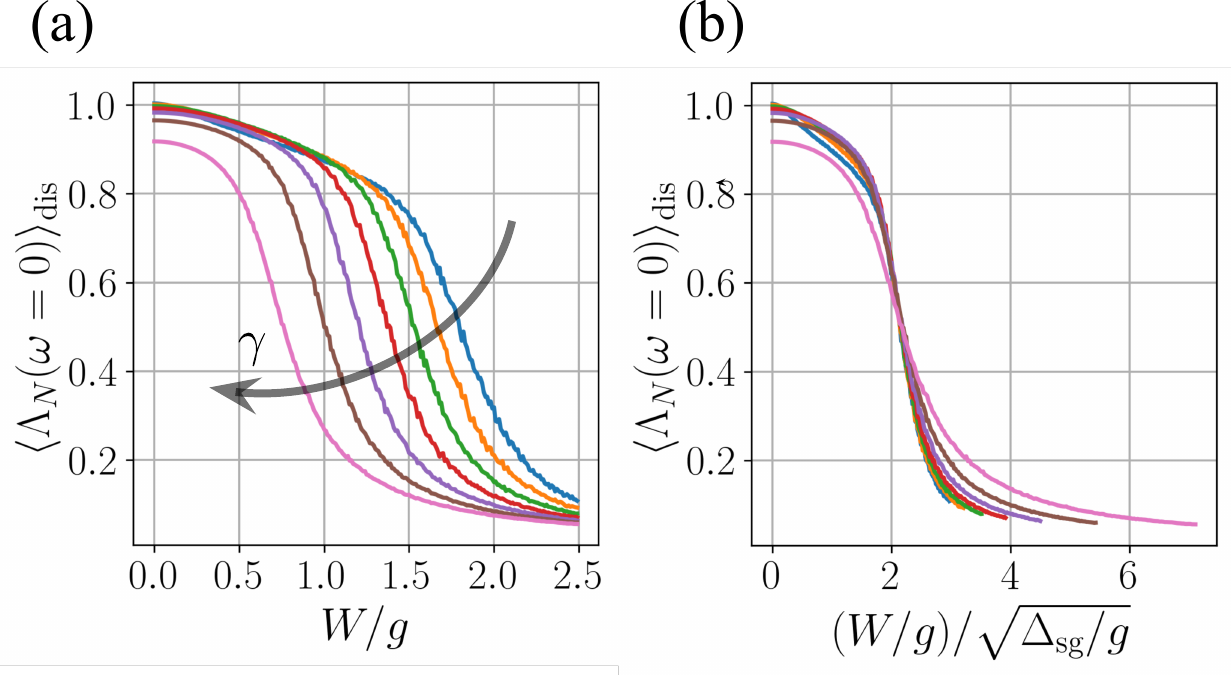}
    \caption{
    Averaged LRO parameter
    with different disorder strengths $W$ at the symmetric parameter regime of Model I, with 
    $\gamma/g = 4.6, 4.8, 5, 5.2, 5.4, 5.6$ (values increasing from blue to magenta). The system size is $N = 100$ and the number of disorder realizations $n_r = 500$. Plots (a) and (b) show the dependence with disorder and the scaling with the singular value gap, respectively.}
    \label{LRO_scaling}
\end{figure}

We focus on the case of local disorder in the on-site energies $\Delta_j$, drawn from a Gaussian distribution with zero mean and standard deviation $W$. We calculate correlation functions averaged over $n_r$ disorder realizations. 
Fig. \ref{LRO_scaling} (a) shows the evolution of the normal LRO at the symmetric point $g_c = g_s = J = g$ for different values of $\gamma$ with $n_r = 500$ realizations, such that we represent the averaged LRO parameter $\bar{\Lambda}_N(\omega)$. The singular value gap grows as $\gamma$ decreases, enhancing the system’s resilience against disorder.
In Fig. \ref{LRO_scaling} (b), we observe that all curves collapse if we rescale disorder with the gap $\Delta_{\rm sg}$ as $W \to W/\sqrt{\Delta_{\rm sg}}$. We compute the point of maximum slope as an estimate of the critical disorder, $W_{\rm c}$, leading to the scaling
\begin{equation}
W_{\rm c}/g = c  \sqrt{\Delta_{\rm sg}/g},
\end{equation}
with $c \approx 2.2$. This scaling can be understood qualitatively by a simple perturbative argument. 
At low disorder, singular values $\mathbb{S}_n$ are corrected by the disorder perturbation. The first, linear order, vanishes after disorder averaging, whereas the second order correction, $\delta \mathbb{S}_n \propto W^2$. Since topological order dissappears for a critical disorder that induces changes of the order of the singular gap, this leads to the estimation 
$\delta \mathbb{S}_n \approx \Delta_{\rm sg} \propto W_{\rm c}^2$.

\begin{figure}[h]
    \centering
    \includegraphics[width=0.7\textwidth]{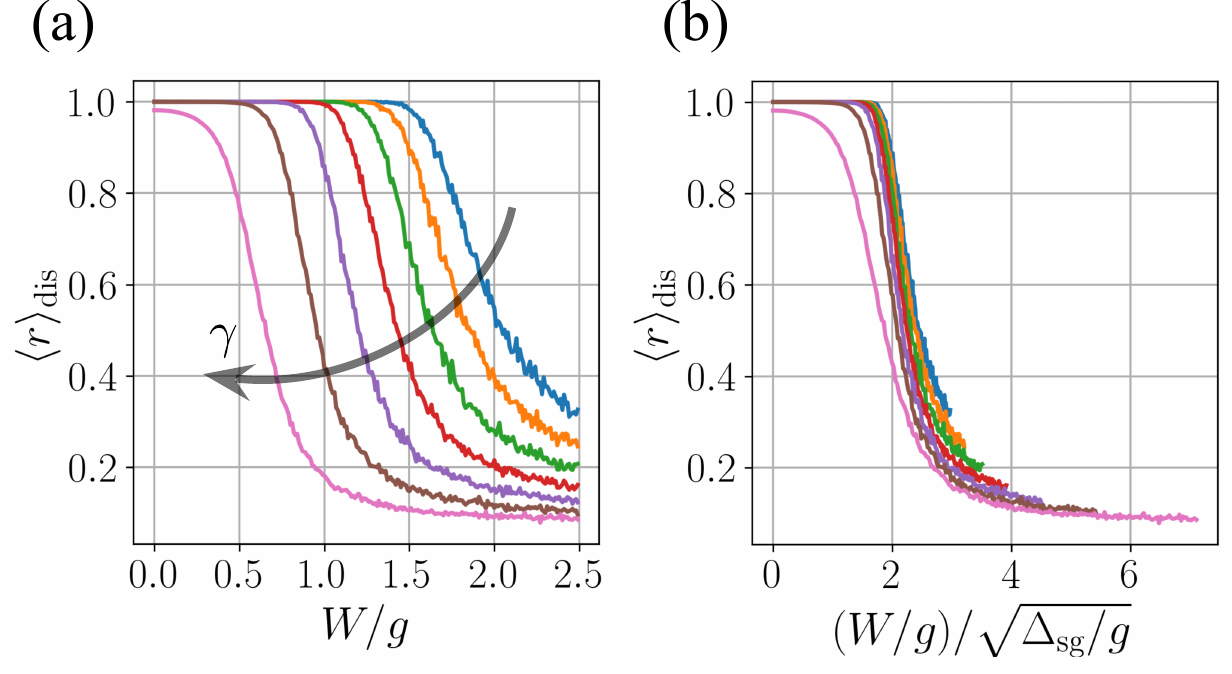}
    \caption{Average over disorder realizations of $r$-parameter at the symmetric point of Model I, and values $\gamma/g = 4.6, 4.8, 5.0, 5.2, 5.4, 5.6, 5.8$, from blue to magenta. The system size is , $N = 100$, and the number of disorder realizations $n_r = 500$. Plots (a) and (b) show the dependence with disorder and the scaling with the singular value gap, respectively.}
    \label{r_scaling}
\end{figure}

To characterize the gap closure, we define the singular-value spacing ratio
\begin{equation}
r(\gamma) \equiv \frac{|\mathbb{S}_1 - \mathbb{S}_0|}{\mathbb{S}_1 + \mathbb{S}_0}, \qquad r(\gamma) \in [0,1],
\end{equation}
where $\mathbb{S}_0$ and $\mathbb{S}_1$ denote the two smallest singular values. In the topological phase, the zero singular value scales as $\mathbb{S}_0 \sim e^{-N/\xi }$, implying $r \to 1$ in the thermodynamic limit. In contrast, in the trivial phase, the two lowest singular 
values become comparable, $\mathbb{S}_0 \approx \mathbb{S}_1$, yielding $r \approx 0$. Fig. \ref{r_scaling} (a) shows the evolution of the $r$-parameter at the symmetric point $g_c = g_s = J = g$ for different values of $\gamma$ with $n_r = 500$ realizations, such that we represent the averaged $r$-parameter. In Fig. \ref{r_scaling} (b), we again observe that all curves collapse if we rescale disorder as $W \to W/\sqrt{\Delta_{\rm sg}}$.
%


%
\begin{figure}[h]
    \centering
    \includegraphics[width=0.75\textwidth]{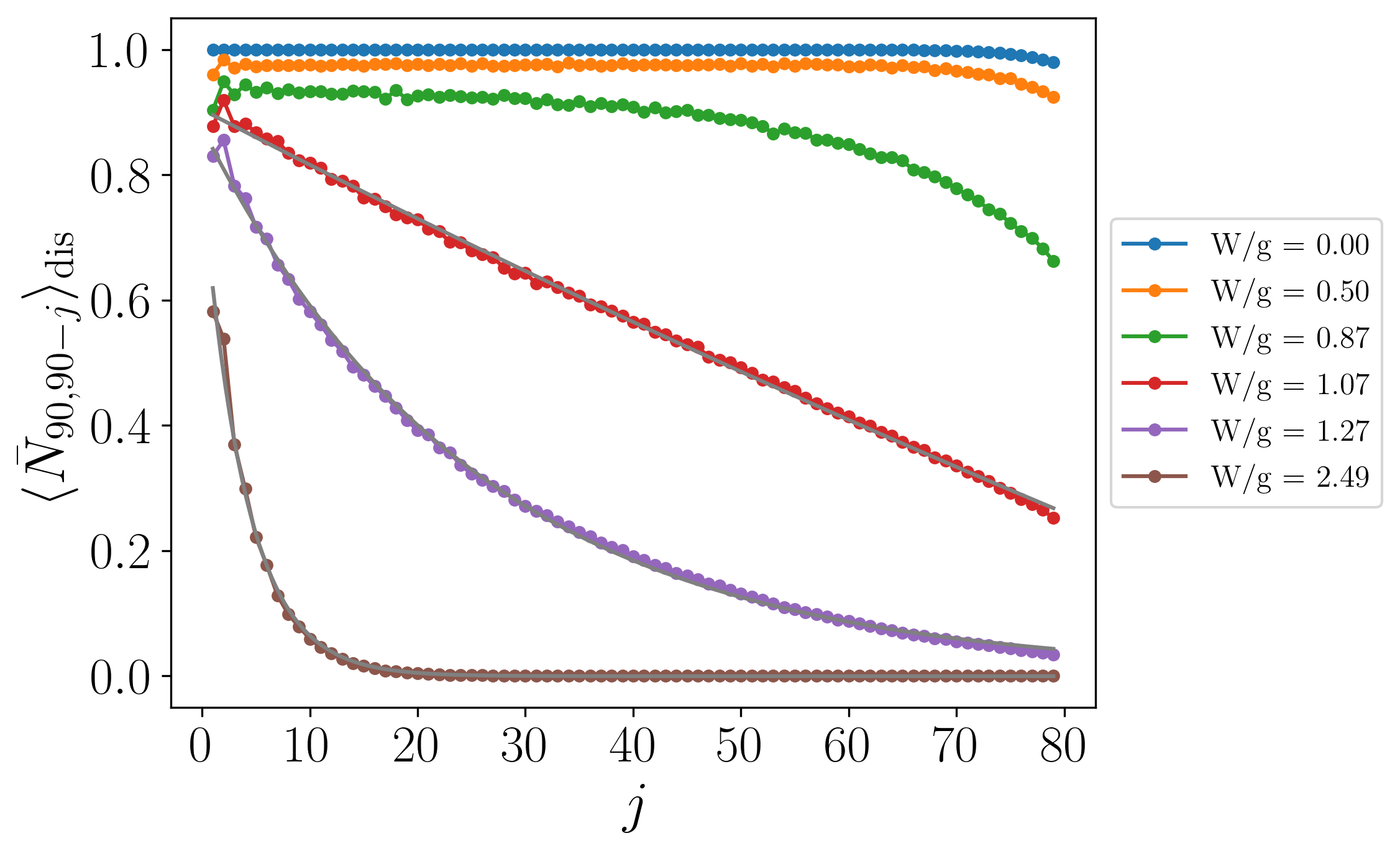}
    \caption{Normalized correlations in frequency space at $\omega = 0$ for different values of the disorder strength at the symmetric regime of Model I, for $\gamma/g = 5.5$. The system size is $N = 100$, and the number of disorder realizations $n_r = 500$. Continuous lines corresponds to exponential fits.}
    \label{noisev}
\end{figure}

We also investigate the spatial decay of normalized frequency-resolved correlations numerically as a function of disorder. In Fig. \ref{noisev}, we show numerical results at the symmetric point with $\gamma/g = 5$ and $n_r = 500$ realizations. Normalized correlations can be classified into two distinct regimes:
\begin{itemize}
\item Low-disorder regime $(W < W_{\rm c})$.- Slow non-standard decay of correlations, following a convex curve that does not follow a usual power-law or exponential decay form. 

\item Disordered phase $(W\geq W_{\rm c})$.- Correlations 
show an exponential decay with a decay length that increases as the system approaches the critical disorder 
$\bar{N}_{ij}(\omega) \propto e^{-j/\xi}$. Exactly at the critical disorder $W  = W_{\rm c}$, correlations show a linear decay that cannot be distinguished from an exponential dependence in the numerical calculations.
\end{itemize}

\begin{figure}[h]
    \centering
    \includegraphics[width=0.6\textwidth]{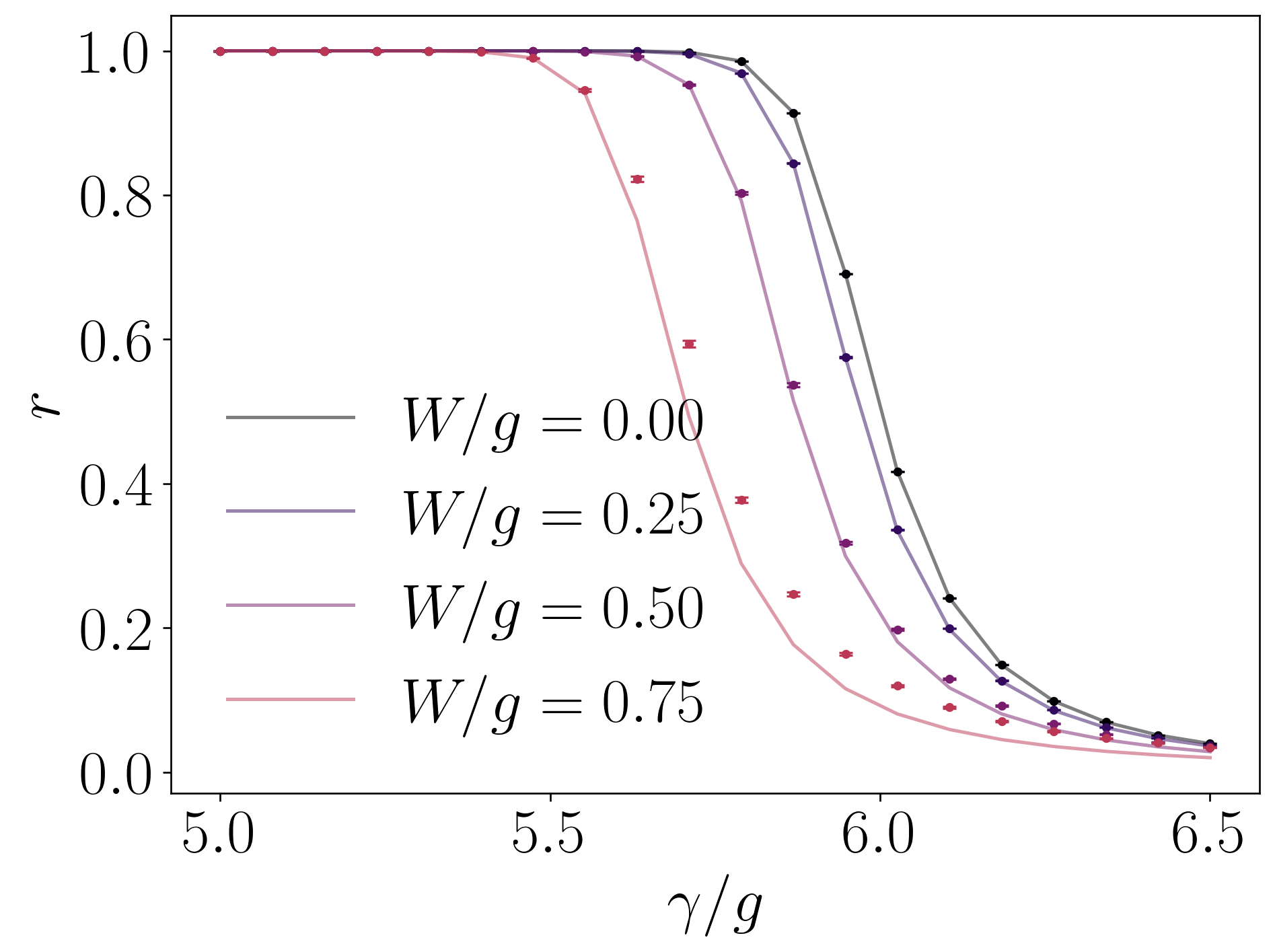}
    \caption{Average over disorder realizations of $r$-parameter as a function of the local dissipation $\gamma/g$ for different values of weak noise $W/g=0.00,0.25,0.50,0.75$. The points represent the results with noise obtained after averaging over 1000 iterations. The solid lines represent the results without noise for the renormalized effective parameters in Eq. (\ref{renmain}). For weak disorder, the renormalized theoretical approximation match with the simulated noisy data. The parameters correspond to the symmetric point defined in Sec.~\ref{modelI} with system size $N=100$.}
    \label{renno}
\end{figure}

To gain further insight, in the limit of small noise $\rm W \ll \rm W_c$, we can treat the weak disorder as a perturbation of the bare Green's function $\mathbb{G}$ (see Appendix \ref{Renorm}), leading to the following renormalized effective parameters in the topological phase within the symmetric parameter regime 
\begin{eqnarray}\label{renmain}
    \Delta_{\rm eff}&\approx&\Delta + W^2\Re{(G_{ii})}, \nonumber \\
    \gamma_{\rm eff}&\approx&\gamma -2 W^2\Im{(G_{ii})}, \nonumber \\
    g_{s_{\rm eff}}&\approx& g_s-W^2(\Bar{G}_{ii}).
\end{eqnarray}
We validate these approximations by analyzing the properties of the singular-value gap associated with the renormalized Green's function. In Fig.~\ref{renno}, we plot the renormalized $r$-parameter $r(\Delta_{\rm eff},\gamma_{\rm eff}, g_{s_{ \rm eff}})$ in absence of noise, as solid lines. The data points, averaged over 1000 realizations for different noise strengths, exhibit excellent agreement with the theoretical prediction in the regime of weak disorder.

\subsection{Topological two-point equal-time correlations}

To characterize equal-time correlations, we integrate the previous frequency-resolved correlation functions over the full frequency spectrum, according to Eq. (\ref{freq.integrated}). Below, we analytically derive the spatial decay of the integrated correlations, showing an enhancement of its spatial extent in the topological phase.

In the topological regime, the frequency-resolved correlations are dominated by the smallest singular value. We consider the case where a single singular value is significantly smaller than the remaining ones. This enables an approximate description based on Eq. (\ref{zerosing}), yielding the following expression for the equal-time correlation between sites $l$ and $j$,
\begin{equation}\label{ap}
\mathcal{N}_{lj} \approx \frac{\gamma}{2\pi}\int^{+\infty}_{-\infty} d\omega\, 
\VV_{l0}^*(\omega)\,\mathbb{\Sigma}_{00}(\omega)\,\VV_{j0}(\omega).
\end{equation}
We now assume that the singular vectors are characterized by an exponentially decaying real envelope together with an oscillatory phase, in agreement with non-Bloch theory \cite{yao2018edge,yokomizo2019non},
\begin{equation}\label{V0}
    \VV_{l0}(\omega)=A(\omega)\, e^{i\Tilde{k}^{+}(\omega)l+\lambda^{+}(\omega)l}.
\end{equation}
 For the symmetric parameter choice $J=g_s=g_c=g$, $\phi=\pi/2$, and $\Delta=0$, we obtain analytical expressions for the amplitude $A(\omega)$, the generalized Bloch momentum $\Tilde{k}^{+}(\omega)$, and the inverse localization length $\lambda^{+}(\omega)$. These quantities are derived in App. \ref{sv} by solving the zero–energy edge-state equation in the semi-infinite chain limit.

One can numerically observe that the product between the singular vector amplitude and the amplification term can be approximated to a frequency independent constant, given by
\begin{equation}
    |A(\omega)|^2\mathbb{\Sigma}_{00}(\omega) \approx\frac{1}{16}\frac{e^{2\lambda^{+}(\omega)}}{1-e^{-2\lambda^{+}(\omega)}}\approx c.
\end{equation}

Replacing the previous expressions into Eq.~(\ref{ap}), the equal-time correlations in the thermodynamic limit $(N\rightarrow\infty)$ are reduced to
\begin{equation}\label{ap2}
\mathcal{N}_{lj} \approx \frac{\gamma c}{2\pi}\int d\omega\, 
e^{-i\Tilde{k}^{+}(\omega)(l-j)+\lambda^{+}(\omega)(l+j)}.
\end{equation}
To make further analytical progress, we expand the phase function $\Tilde{k}^{+}(\omega)$, derived in App. \ref{sv}, to linear order in frequency. For $|\omega|\ll \gamma-2$, this yields
\begin{eqnarray}
    \Tilde{k}^{+}(\omega)\approx \frac{2}{\gamma-2}\,\omega.
\end{eqnarray}
Similarly, expanding $\lambda^{+}(\omega)$ up to second order in $\omega$, we obtain
\begin{eqnarray}
    \lambda^{+}(\omega)\approx \lambda_0-\frac{2\omega^2}{(\gamma-2)^2},
\qquad
\lambda_0=\ln\frac{4}{\gamma-2}.
\end{eqnarray}
Inserting both approximations into Eq.~(\ref{ap2}) and performing the Gaussian integration, we find
\begin{equation}
\mathcal{N}_{lj}
\approx
\gamma c\frac{(\gamma-2)}{2\pi}\,
e^{\lambda_0(l+j)}
\sqrt{\frac{\pi}{2(l+j)}}
e^{-\frac{(l-j)^2}{2(l+j)}}.
\end{equation}
The corresponding normalized equal-time correlations in the topological phase take the simple form
\begin{equation}\label{prediction}
\Bar{\mathcal{N}}_{ij}
\approx
\sqrt{\frac{2\sqrt{lj}}{l+j}}
\;
e^{-\frac{(l-j)^2}{2(l+j)}}.
\end{equation}
Equivalently, for the anomalous integrated correlations, we obtain $\Bar{\mathcal{M}}_{ij}=i\,\Bar{\mathcal{N}}_{ij}.$ 

\begin{figure}[h]
    \centering
    \includegraphics[width=1\textwidth]{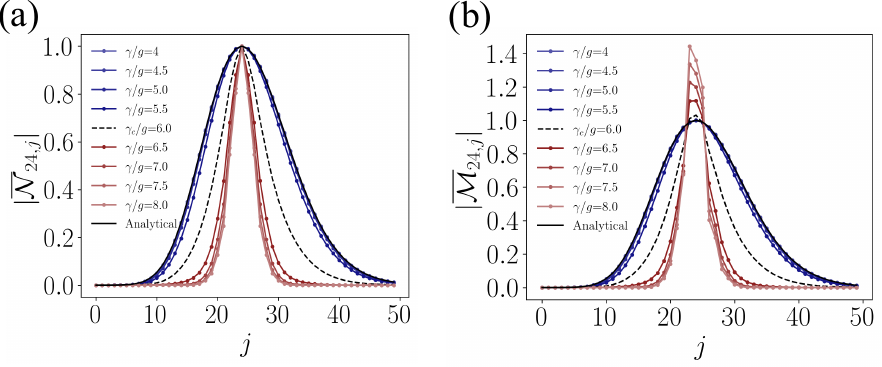}
    \caption{Model I: Normalized equal-time normal (a), and anomalous (b) correlations centered at $i=24$ as a function of the distance $j$ for various dissipation values $\gamma/g$. The solid black line shows the theoretical prediction of Eq. (\ref{prediction}). The critical value $\gamma_c/g=6$ (dashed black line) represents the interface between topological ($\Vec{\mathcal{V}_1}=(0,1,0)$) and trivial phase ($\Vec{\mathcal{V}_0}=(0)$). The parameters correspond to the symmetric point defined in Sec.~\ref{modelI}, with $N=50$.}
    \label{Nnorm}
\end{figure}

While in Sec.~\ref{6.1} we showed that frequency-resolved correlations exhibit LRO in the topological phase, the corresponding equal-time correlations instead show a Gaussian-like spatial decay with variance $\sigma^2 = l + j$. This behavior arises from destructive interference among different frequency components and accounts for the suppression of long-range order in the time domain, even deep within the topological regime.
The resulting normalized correlations for Model~I are presented in Fig.~\ref{Nnorm}. The numerical results (blue dots), corresponding to the topological phase $\Vec{\mathcal{V}_1} = (0,1,0)$, show excellent agreement with the theoretical prediction of Eq.~(\ref{prediction}) (solid black line), indicating that within the topological phase the Gaussian width is independent of the free parameter $\gamma$. Although these correlations do not exhibit long-range order, their spatial decay is slower than the purely exponential decay characteristic of trivial configurations (red dots), corresponding to the trivial phase $\Vec{\mathcal{V}_0} = (0)$.

\begin{figure}[h]
    \centering
    \includegraphics[width=1\textwidth]{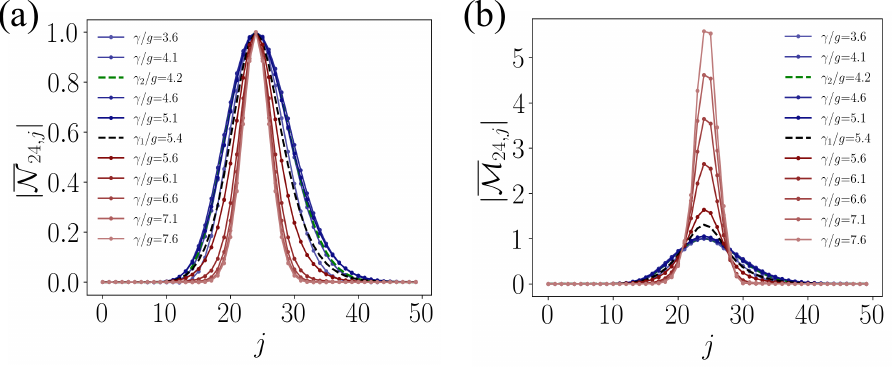}
    \caption{Model II: Normalized equal-time normal (a), and anomalous (b) correlations centered at $i=24$ as a function of the distance $j$ for various dissipation values $\gamma/g$. The critical value $\gamma_{c_1}/g=5.4$ (green dashed line) represents the interface between topological ($\Vec{\mathcal{V}_1}=(0,1,0)$) and trivial phase ($\Vec{\mathcal{V}_0}=(0)$). There is another critical value between two topological phases from ($\Vec{\mathcal{V}_2}=(0,1,2,1,0)$) to ($\Vec{\mathcal{V}_1}=(0,1,0)$)   at $\gamma_{c_2}/g=4.2$ (black dashed line). The remaining parameters are defined in Sec. \ref{modelII}, with $N=50$, $g_c'/g=3$, and $\gamma'/g=30$ (corresponding to $P/g=\frac{3}{5}$).}
    \label{N224j}
\end{figure}

The previous analytical discussion applies to Model~I. Although Model~II allows for winding numbers larger than one, one singular value remains more suppressed than the others (see Fig.~\ref{s1s2} in App.~\ref{sv}). As a consequence, Fig.~\ref{N224j} show a qualitatively similar Gaussian-like behavior of the correlations in Model~II. More specifically, these figures display the normalized equal-time correlations as a function of the distance $|i-j|$, revealing a clear crossover between two regimes: in the topological phase, correlations decay following a broad Gaussian envelope, whereas in the trivial phase they exhibit a much sharper exponential suppression. This behavior reflects the fact that, despite the presence of multiple singular channels in Model~II, the correlations are effectively dominated by a single mode across most of the parameter range.

In contrast to Model~I, Model~II exhibits two topological critical points: the system transitions from $\Vec{\mathcal{V}_2} = (0,1,2,1,0)$ to $\Vec{\mathcal{V}_1} = (0,1,0)$ at $\gamma_1 = 4.2$, and from $\Vec{\mathcal{V}_1} = (0,1,0)$ to the trivial phase $\Vec{\mathcal{V}_0} = (0)$ at $\gamma_2 = 5.4$. The first transition at $\gamma_1$ modifies the number of topological channels but does not qualitatively affect the spatial structure of correlations, as the dominant singular value remains well separated from the rest. By contrast, the second transition at $\gamma_2$ is associated with the closing of the last singular-value gap and the disappearance of the dominant topological mode. Thus, the relevant critical point is $\gamma_2$, as it corresponds to the transition from a topological to a fully trivial phase. This distinction is directly reflected in the behavior of the correlation profiles shown in Fig.~\ref{N224j}: while no qualitative change is observed across $\gamma_1$, a clear transition from Gaussian-like to exponential decay takes place at $\gamma_2$. In particular, the width of the Gaussian envelope decreases as the system approaches $\gamma_2$ from below, signaling the progressive loss of long-range coherence.

\begin{figure}[h]
    \centering
    \includegraphics[width=0.85\textwidth]{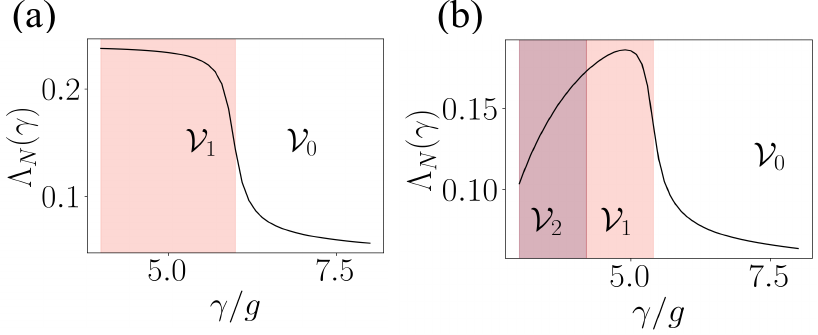}
    \caption{LRO parameter as a function of the dissipation rate $\gamma/g$ for  normal ($\Lambda_\mathcal{N}$, solid black line) and anomalous ($\Lambda_\mathcal{M}$, dashed red line) equal-time correlations. (a) Model I: a critical point at $\gamma_c/g = 6$ signals the transition from the topological phase $\Vec{\mathcal{V}}_1 = (0,1,0)$ to the trivial phase $\Vec{\mathcal{V}}_0 = (0)$, for the symmetric parameter set defined in Sec.~\ref{modelI}. (b) Model II: two critical point are observed, at $\gamma_{c_1}/g = 4.2$, marking the transition from $\Vec{\mathcal{V}}_1 = (0,1,2,1,0)$ to $\Vec{\mathcal{V}}_2 = (0,1,0)$, and at $\gamma_{c_2}/g = 5.4$, signaling the transition to the trivial phase.  The remaining parameters are defined in Sec. \ref{modelII} with $\gamma/g=3.75$, $g_c'/g=3$, and $\gamma'/g=30$ (corresponding to $P/g=\frac{3}{5}$). The system size for both models is $N=80$.}
    \label{MODELO1}
\end{figure}

In particular, Fig.~\ref{MODELO1} (a) and (b) show that, for both Models~I and II, the LRO parameter exhibits a transition at the boundary between $\Vec{\mathcal{V}_1} = (0,1,0)$ and $\Vec{\mathcal{V}_0} = (0)$. These figures display the global LRO parameter as a function of the control parameter $\gamma$, revealing a sharp drop at the critical point. In the topological phase, the LRO parameter remains finite, reflecting the persistence of non-decaying correlations in frequency space, whereas it rapidly approaches zero in the trivial regime.

\newpage
\section{Conclusions and outlook}\label{section9}
In summary, we have developed a comprehensive framework to characterize the impact of non-Hermitian topological phases on correlations in driven–dissipative bosonic chains, explicitly distinguishing between frequency-resolved and equal-time observables. By extending the notion of adiabatic deformation from closed Hamiltonian systems to quadratic Liouvillians, we introduced a frequency-resolved topological characterization in terms of a winding-number array defined over the entire frequency domain.

Our results establish three central findings. First, we uncover a general mechanism linking the topology of the singular-value spectrum to the emergence of long-range order in frequency-resolved correlations of open bosonic systems. Second, this long-range order is robust against disorder, provided that the dissipative gap remains open. Third, equal-time correlations in the topological phase display a qualitatively distinct spatial structure, characterized by an enhanced Gaussian decay rather than the conventional exponential behavior found in trivial phases.

The origin of long-range order can be traced to the effective rank-one structure of the frequency-resolved correlation matrix, which becomes dominated by a single singular mode. In this regime, fluctuations across the system are governed by a single collective degree of freedom. From a physical standpoint, topological amplification selectively channels quantum noise into a global mode, leading to phase coherence across distant sites. This mechanism establishes a direct analogy with phenomena such as Bose–Einstein condensation and phase locking in lasers, but realized here in a purely driven–dissipative and topological setting.

In contrast, equal-time correlations exhibit an extended spatial profile within the topological phase, with a Gaussian decay emerging from the interference of contributions across different frequencies. This highlights the nontrivial interplay between spectral and spatial structures in open quantum systems.

This work
opens several research directions, including the behavior of correlations in nonlinear and non-Markovian topological systems. Although some approaches based on the Schwinger-Keldysh formalism have already been developed for the Bose-Hubbard model, showing exponential decay of correlations in interacting systems \cite{grass2019excitations}, a compelling direction would be to explore the emergence of long-range order in the presence of interactions within a topological phase.
Furthermore, next works will focus on distinguishing quantum correlations within the previous topological models in order to understand the crossover between topological phases and entanglement \cite{roccati2023quantum}.

\ack{This work is supported by the Spanish projects PID2021-127968NBI00 and PID2024-
159152NB-I00 (D. P. and M. C. R.), and by
PID2023-146531NA-I00 (T. R.), financed by MCIN/AEI/10.13039/501100011033 and
ERDF/EU. We also acknowledge support from CAM Programa TEC-2024/COM-84 QUITEMAD-CM.
T. R. further acknowledges the Ram\'on
y Cajal program RYC2021-032473-I, financed by
MCIN/AEI/10.13039/501100011033 and the European Union NextGenerationEU/PRTR.}
\data{All data that support the findings of this study are included within the article.}

\newpage

\bibliographystyle{iopart-num}
\bibliography{topbib}

@article{brunelli2023restoration,
  title={Restoration of the non-Hermitian bulk-boundary correspondence via topological amplification},
  author={Brunelli, Matteo and Wanjura, Clara C and Nunnenkamp, Andreas},
  journal={SciPost Physics},
  volume={15},
  number={4},
doi={10.21468/SciPostPhys.15.4.173},
  pages={173},
  year={2023}
}

@article{moustaj2022field,
  title={Field theoretical study of disorder in non-Hermitian topological models},
  author={Moustaj, Anouar and Eek, Lumen and Morais Smith, Cristiane},
  journal={Phys. Rev. B},
  volume={105},
  number={18},
doi={https://doi.org/10.1103/PhysRevB.105.L180503},
  pages={L180503},
  year={2022},
  publisher={APS}
}

@article{roccati2023quantum,
  title={Quantum correlations in dissipative gain--loss systems across exceptional points},
  author={Roccati, Federico and Purkayastha, Archak and Palma, G Massimo and Ciccarello, Francesco},
  journal={Eur. Phys. J. Spec. Top.},
  volume={232},
  number={11},
  pages={1783--1788},
  year={2023},
doi={https://doi.org/10.1140/epjs/s11734-023-00835-3},
  publisher={Springer}
}

@article{feinberg1997non,
  title={Non-hermitian random matrix theory: Method of hermitian reduction},
  author={Feinberg, Joshua and Zee, Anthony},
  journal={Nucl. Phys. B},
  volume={504},
  number={3},
  doi={https://doi.org/10.1016/S0550-3213(97)00502-6},
  pages={579--608},
  year={1997},
  publisher={Elsevier}
}

@article{rassaert2025emerging,
  title={Emerging Non-Hermitian Topology in a Chiral-Driven-Dissipative Bose-Hubbard Model},
  author={Rassaert, Laszlo and Ramos, Tom{\'a}s and Roscilde, Tommaso and Porras, Diego},
  journal={Phys. Rev. Lett.},
  volume={135},
  number={20},
  pages={203603},
  doi={10.1103/w32k-x3l5},
  year={2025},
  publisher={APS}
}

@article{Ramos:2022kvf,
    author = "Ramos, Tom{\'a}s and G{\'o}mez-Le{\'o}n, {\'A}lvaro and Garc{\'\i}a-Ripoll, Juan Jos{\'e} and Gonz{\'a}lez-Tudela, Alejandro and Porras, Diego",
    title = "{Topological Josephson parametric amplifier array: A proposal for directional, broadband, and low-noise amplification}",
    eprint = "2207.13728",
    archivePrefix = "arXiv",
    primaryClass = "quant-ph",
    month = "7",
doi={
https://doi.org/10.48550/arXiv.2207.13728
},
    year = "2022"
}

@article{yokomizo2019non,
  title={Non-Bloch band theory of non-Hermitian systems},
  author={Yokomizo, Kazuki and Murakami, Shuichi},
  journal={Phys. Rev. Lett.},
  volume={123},
  number={6},
doi={ https://doi.org/10.1103/PhysRevLett.123.066404},
  pages={066404},
  year={2019},
  publisher={APS}
}

@article{parra2025floquet,
  title={Floquet Topological Frequency-Converting Amplifier},
  author={Parra-Rodriguez, Adrian and Clavero-Rubio, Miguel and Gigon, Philippe and Ramos, Tom{\'a}s and G{\'o}mez-Le{\'o}n, {\'A}lvaro and Porras, Diego},
  year={2025},
  archivePrefix={arXiv},
      eprint={2512.08880},      
      doi={10.48550/arXiv.2512.08880}
}

@article{pernet2022gap,
  title={Gap solitons in a one-dimensional driven-dissipative topological lattice},
  author={Pernet, Nicolas and St-Jean, Philippe and Solnyshkov, Dmitry D and Malpuech, Guillaume and Carlon Zambon, Nicola and Fontaine, Quentin and Real, Bastian and Jamadi, Omar and Lema{\^\i}tre, Aristide and Morassi, Martina and others},
  journal={Nat. Phys.},
  volume={18},
doi={https://doi.org/10.1038/s41567-022-01599-8},
  number={6},
  pages={678--684},
  year={2022},
  publisher={Nature Publishing Group UK London}
}

@article{bardyn2013topology,
  title={Topology by dissipation},
  author={Bardyn, Charles-Edouard and Baranov, Mikhail A and Kraus, Christina V and Rico, Enrique and {\.I}mamo{\u{g}}lu, A and Zoller, Peter and Diehl, Sebastian},
  journal={New J. Phys.},
  volume={15},
  number={8},
  pages={085001},
doi={10.1088/1367-2630/15/8/085001},
year={2013},
  publisher={IOP Publishing}
}

@article{villa2025topological,
  title={Topological classification of driven-dissipative nonlinear systems},
  author={Villa, Greta and Del Pino, Javier and Dumont, Vincent and Rastelli, Gianluca and Micha{\l}ek, Mateusz and Eichler, Alexander and Zilberberg, Oded},
  journal={Sci. Adv.},
  volume={11},
  number={33},
  pages={eadt9311},
  year={2025},
doi={10.1126/sciadv.adt9311},
  publisher={American Association for the Advancement of Science}
}

@article{herviou2019defining,
  title={Defining a bulk-edge correspondence for non-Hermitian Hamiltonians via singular-value decomposition},
  author={Herviou, Loic and Bardarson, Jens H and Regnault, Nicolas},
  journal={Phys. Rev. A},
  volume={99},
  number={5},
doi={https://doi.org/10.1103/PhysRevA.99.052118},
  pages={052118},
  year={2019},
  publisher={APS}
}

@article{yao2018edge,
  title={Edge states and topological invariants of non-Hermitian systems},
  author={Yao, Shunyu and Wang, Zhong},
  journal={Phys. Rev. Lett.},
  volume={121},
  number={8},
  pages={086803},
  doi={https://doi.org/10.1103/PhysRevLett.121.086803},
  year={2018},
  publisher={APS}
}

@misc{okuma2026steadystateskineffectbosonic,
      title={Steady-state skin effect in bosonic topological edge states under parametric driving}, 
      author={Nobuyuki Okuma},
      year={2026},
      journal={arXiv:2602.01625},     
      eprint={2602.01625},
      archivePrefix={arXiv},
      doi={10.1103/xvp5-ptqq}, 
}

@article{hasan2010colloquium,
  title={Colloquium: topological insulators},
  author={Hasan, M Zahid and Kane, Charles L},
  journal={Rev. Mod. Phys.},
  volume={82},
  number={4},
  pages={3045--3067},
  year={2010},
doi={https://doi.org/10.1103/RevModPhys.82.3045},
  publisher={APS}
}

@article{bansil2016colloquium,
  title={Colloquium: Topological band theory},
  author={Bansil, Arun and Lin, Hsin and Das, Tanmoy},
  journal={Rev. Mod. Phys.},
  volume={88},
  number={2},
  pages={021004},
  year={2016},
doi={https://doi.org/10.1103/RevModPhys.88.021004},
  publisher={APS}
}

@article{kitaev2003fault,
  author  = {Kitaev, A. Yu.},
  title   = {Fault-tolerant quantum computation by anyons},
  journal = {Ann. Phys.},
  volume  = {303},
  pages   = {2--30},
  year    = {2003},
  doi     = {10.1016/S0003-4916(02)00018-0}
}

@article{koch2022quantum,
  author  = {Koch, F. and Budich, J. C.},
  title   = {Quantum non-Hermitian topological sensors},
  journal = {Phys. Rev. Res.},
  volume  = {4},
  pages   = {013113},
  year    = {2022},
  doi     = {10.1103/PhysRevResearch.4.013113}
}

@article{slim2024optomechanical,
  author  = {Slim, J. J. and Wanjura, C. C. and Brunelli, M. and Del Pino, J. and Nunnenkamp, A. and Verhagen, E.},
  title   = {Optomechanical realization of the bosonic Kitaev chain},
  journal = {Nature},
  volume  = {627},
  pages   = {767--771},
  year    = {2024},
  doi     = {10.1038/s41586-024-07174-w}
}

@article{klitzing1980new,
  author  = {von Klitzing, K. and Dorda, G. and Pepper, M.},
  title   = {New method for high-accuracy determination of the fine-structure constant based on quantized Hall resistance},
  journal = {Phys. Rev. Lett.},
  volume  = {45},
  pages   = {494},
  year    = {1980},
  doi     = {10.1103/PhysRevLett.45.494}
}

@article{ashida2020non,
  title={Non-hermitian physics},
  author={Ashida, Yuto and Gong, Zongping and Ueda, Masahito},
  journal={Adv. Phys.},
  volume={69},
  number={3},
  pages={249--435},
  year={2020},
doi={https://doi.org/10.1080/00018732.2021.1876991},
  publisher={Taylor \& Francis}
}

@misc{gu2026nonreciprocityenrichedsteadyphasesopen,
      title={Nonreciprocity-enriched steady phases in open quantum systems}, 
      author={Ding Gu and Zhanpeng Fu and Zhong Wang},
      year={2026},
      journal={arxiv:2605.00101},
      eprint={2605.00101},
      archivePrefix={arXiv},
      doi={10.48550/arXiv.2605.00101}, 
}

@article{dalfovo1999theory,
  title={Theory of Bose-Einstein condensation in trapped gases},
  author={Dalfovo, Franco and Giorgini, Stefano and Pitaevskii, Lev P and Stringari, Sandro},
  journal={Reviews of modern physics},
  volume={71},
  number={3},
  pages={463},
  year={1999},
  doi={https://doi.org/10.1103/RevModPhys.71.463},
  publisher={APS}
}

@article{okuma2023non,
  title={Non-Hermitian topological phenomena: A review},
  author={Okuma, Nobuyuki and Sato, Masatoshi},
  journal={Annu. Rev. Condens. Matter Phys.},
  volume={14},
  number={1},
doi={https://doi.org/10.1146/annurev-conmatphys-040521-033133},
  pages={83--107},
  year={2023},
  publisher={Annual Reviews}
}

@article{ramos2021topological,
  title={Topological input-output theory for directional amplification},
  author={Ramos, Tom{\'a}s and Garc{\'\i}a-Ripoll, Juan Jos{\'e} and Porras, Diego},
  journal={Phys. Rev. A},
  volume={103},
  number={3},
  pages={033513},
  year={2021},
doi={https://doi.org/10.1103/PhysRevA.103.033513},
  publisher={APS}
}

@article{asboth2016short,
  title={A short course on topological insulators},
  author={Asb{\'o}th, J{\'a}nos K and Oroszl{\'a}ny, L{\'a}szl{\'o} and P{\'a}lyi, Andr{\'a}s},
  journal={Lecture notes in Physics},
  volume={919},
  number={1},
doi={
https://doi.org/10.48550/arXiv.1509.02295},
  year={2016},
  publisher={Springer}
}

@article{qi2011topological,
  title={Topological insulators and superconductors},
  author={Qi, Xiao-Liang and Zhang, Shou-Cheng},
  journal={Rev. Mod. Phys.},
  volume={83},
  number={4},
  pages={1057},
doi={https://doi.org/10.1103/RevModPhys.83.1057},
  year={2011},
  publisher={APS}
}

@article{porras2019topological,
  author  = {Porras, D. and Fernández-Lorenzo, S.},
  title   = {Topological amplification in photonic lattices},
  journal = {Phys. Rev. Lett.},
  volume  = {122},
  pages   = {143901},
  year    = {2019},
  doi     = {10.1103/PhysRevLett.122.143901}
}

@article{wanjura2020topological,
  author  = {Wanjura, C. C. and Brunelli, M. and Nunnenkamp, A.},
  title   = {Topological framework for directional amplification in driven-dissipative cavity arrays},
  journal = {Nat. Commun.},
  volume  = {11},
  pages   = {3149},
  year    = {2020},
  doi     = {10.1038/s41467-020-16863-9}
}

@article{gomez2023driven,
  author  = {Gomez-Leon, A. and Ramos, T. and Gonzalez-Tudela, A. and Porras, D.},
  title   = {Driven-dissipative topological phases in parametric resonator arrays},
  journal = {Quantum},
  volume  = {7},
  pages   = {1016},
  year    = {2023},
  doi     = {10.22331/q-2023-05-23-1016}
}

@article{grass2019excitations,
  author  = {Graß, T.},
  title   = {Excitations and correlations in the driven-dissipative Bose-Hubbard model},
  journal = {Phys. Rev. A},
  volume  = {99},
  pages   = {043607},
  year    = {2019},
  doi     = {10.1103/PhysRevA.99.043607}
}

@article{peano2016topological,
  author  = {Peano, V. and Houde, M. and Marquardt, F. and Clerk, A. A.},
  title   = {Topological quantum fluctuations and traveling wave amplifiers},
  journal = {Phys. Rev. X},
  volume  = {6},
  pages   = {041026},
  year    = {2016},
  doi     = {10.1103/PhysRevX.6.041026}
}

@article{mcdonald2018phase,
  title={Phase-dependent chiral transport and effective non-Hermitian dynamics in a bosonic Kitaev-Majorana chain},
  author={McDonald, Alexander and Pereg-Barnea, T and Clerk, AA},
  journal={Phys. Rev. X},
  volume={8},
  number={4},
doi={https://doi.org/10.1103/PhysRevX.8.041031},
  pages={041031},
  year={2018},
  publisher={APS}
}

@article{kawabata2019symmetry,
  title={Symmetry and topology in non-Hermitian physics},
  author={Kawabata, Kohei and Shiozaki, Ken and Ueda, Masahito and Sato, Masatoshi},
  journal={Phys. Rev. X},
  volume={9},
  number={4},
doi={https://doi.org/10.1103/PhysRevX.9.041015},
  pages={041015},
  year={2019},
  publisher={APS}
}

@article{okuma2020topological,
  title={Topological origin of non-Hermitian skin effects},
  author={Okuma, Nobuyuki and Kawabata, Kohei and Shiozaki, Ken and Sato, Masatoshi},
doi={https://doi.org/10.1103/PhysRevLett.124.086801},
  journal={Phys. Rev. Lett.},
  volume={124},
  number={8},
  pages={086801},
  year={2020},
  publisher={APS}
}

@misc{ughrelidze2026quantumcriticalitythermodynamicstability,
      title={Quantum criticality beyond thermodynamic stability}, 
      author={Mariam Ughrelidze and Vincent P. Flynn and Emilio Cobanera and Lorenza Viola},
      year={2026},
      eprint={2605.04153},
      archivePrefix={arXiv},
      doi={arXiv.2605.04153
} 
}

@article{Flynn21prl,
  title = {Topology by Dissipation: Majorana Bosons in Metastable Quadratic Markovian Dynamics},
  author = {Flynn, Vincent P. and Cobanera, Emilio and Viola, Lorenza},
  journal = {Phys. Rev. Lett.},
  volume = {127},
  pages = {245701},
  numpages = {7},
  year = {2021},
  month = {Dec},
  publisher = {American Physical Society},
  doi = {10.1103/PhysRevLett.127.245701},
}

@article{Flynn23prb,
  title = {Topological zero modes and edge symmetries of metastable Markovian bosonic systems},
  author = {Flynn, Vincent P. and Cobanera, Emilio and Viola, Lorenza},
  journal = {Phys. Rev. B},
  volume = {108},
  pages = {214312},
  numpages = {35},
  year = {2023},
  month = {Dec},
  publisher = {American Physical Society},
  doi = {10.1103/PhysRevB.108.214312},
}

@article{gomez2022bridging,
  title={Bridging the gap between topological non-Hermitian physics and open quantum systems},
  author={G{\'o}mez-Le{\'o}n, {\'A}lvaro and Ramos, Tom{\'a}s and Gonz{\'a}lez-Tudela, Alejandro and Porras, Diego},
  journal={Phys. Rev. A},
  volume={106},
  number={1},
doi={https://doi.org/10.1103/PhysRevA.106.L011501},
  pages={L011501},
  year={2022},
  publisher={APS}
}

@article{clavero2025vibrational,
  title = {Vibrational parametric arrays with trapped ions: Non-Hermitian topological phases and quantum sensing},
  author = {Clavero-Rubio, Miguel and Ramos, Tom\'as and Porras, Diego},
  journal = {Phys. Rev. Res.},
  volume = {7},
  pages = {043218},
  numpages = {17},
  year = {2025},
  month = {Nov},
  publisher = {American Physical Society},
  doi = {10.1103/jrv5-hmt8}
}

@article{mcdonald2022prb,
  title = {Nonequilibrium stationary states of quantum non-Hermitian lattice models},
  author = {McDonald, A. and Hanai, R. and Clerk, A. A.},
  journal = {Phys. Rev. B},
  volume = {105},
  pages = {064302},
  numpages = {19},
  year = {2022},
  month = {Feb},
  publisher = {American Physical Society},
  doi = {10.1103/PhysRevB.105.064302}
}

@book{horn2012matrix,
  title     = {Matrix Analysis},
  author    = {Horn, Roger A. and Johnson, Charles R.},
  year      = {2012},
  edition   = {2},
  publisher = {Cambridge University Press},
  address   = {Cambridge},
  isbn      = {9780521548236},
  nolink = {}
}

\begin{appendices}

\section{Symmetry constraints on the dynamical matrix} \label{ph}
\subsection{Particle-hole symmetry of the dynamical matrix}\label{phsdm}
The non-Hermitian dynamical matrix $\mathbb{H}$ has a remarkable particle-hole symmetry, defined in terms of the antiunitary symmetry operator $\Xi = C K$, where $C = \mathbb{1}_{N} \otimes\sigma_x$, and $K$ is complex conjugation,
\begin{equation}
\Xi  \mathbb{H} \Xi  = C \mathbb{H}^* C = - \mathbb{H} .
\end{equation}
This symmetry is independent of the particular model parameters, and it 
immediately implies the following property of the Green's function 

\begin{equation}
C  \mathbb{G}^*(\omega) C  = - \mathbb{G}(-\omega).
\end{equation}
If we express the Green's function in $N \times N$ blocks,
\begin{equation}
\mathbb{G}(\omega) = 
\begin{pmatrix}
G(\omega) &&\bar{G}(\omega) \\
\bar{G}'(\omega) &&  G'(\omega)
\end{pmatrix},
\label{eq:G.blocks}
\end{equation}
we obtain the following equivalences,
\begin{eqnarray}\label{phs}
G'(\omega) &=& - G^*(-\omega) , \nonumber \\
\bar{G}'(\omega) &=& - \bar{G}^*(-\omega).
\end{eqnarray}

Particle-hole symmetry implies that singular values of $\omega - \mathbb{H}$ are symmetric in frequency space. To prove this result, we write
$\omega - \mathbb{H} = \mathbb{U}_\omega \mathbb{S}_\omega \mathbb{V}^\dagger_\omega$, showing explicitly the $\omega$-dependence in the SVD. Then
\begin{eqnarray}
C \left(\omega - \mathbb{H}^* \right) C  
&=& -(-\omega - \mathbb{H} ) , \nonumber \\
C \mathbb{U}_\omega^* \mathbb{S}_\omega \left( \mathbb{V_\omega}^\dagger\right)^*  C &=&
- \mathbb{U}_{-\omega} \mathbb{S}_{-\omega} \mathbb{V}^\dagger_{-\omega},
\end{eqnarray}
from which we obtain (up to a global phase) 
$\mathbb{U}_{-\omega} = - C \mathbb{U}^*_\omega$, 
$\mathbb{V}_{-\omega} =  C \mathbb{V}^*_\omega$ and, eventually, the symmetry of singular values under reflection in frequency space,
\begin{equation}
\mathbb{S}_{-\omega} = \mathbb{S}_{\omega} .
\end{equation}
\subsection{Symmetric parameter regime}\label{SPR}

We can exploit a parameter--dependent symmetry, namely,
\begin{eqnarray}
\mathcal{Y}\equiv (\mathbb{1}_N\otimes\sigma_y),
\end{eqnarray}
which emerges when setting the usual parameter configuration with $\Delta=0$ and $\phi=\pi/2$. 
In this regime, the dynamical matrix satisfies
\begin{eqnarray}
\mathcal{Y}(\omega\mathbb{1}_{2N}-\mathbb{H})\mathcal{Y}
=(\omega\mathbb{1}_{2N}-\mathbb{H}),
\end{eqnarray}
which directly implies $[\mathcal{Y},(\omega\mathbb{1}_{2N}-\mathbb{H})]=0.$

Let $|\mathbb{V}_\alpha\rangle$ be a right singular vector of 
$(\omega\mathbb{1}_{2N}-\mathbb{H})$ associated with the singular value 
$\mathbb{S}_\alpha$, i.e.
\begin{eqnarray}
(\omega\mathbb{1}_{2N}-\mathbb{H})|\mathbb{V}_\alpha\rangle
=\mathbb{S}_\alpha\,|\mathbb{U}_\alpha\rangle.
\end{eqnarray}
Since $\mathcal{Y}$ commutes with the operator, it follows that
\begin{eqnarray}
(\omega\mathbb{1}_{2N}-\mathbb{H})(\mathcal{Y}|\mathbb{V}_\alpha\rangle)
=\mathbb{S}_\alpha(\mathcal{Y}|\mathbb{U}_\alpha\rangle),
\end{eqnarray}
so that $\mathcal{Y}|\mathbb{V}_\alpha\rangle$ (and analogously 
$\mathcal{Y}|\mathbb{U}_\alpha\rangle$) is also a singular vector associated with 
the same singular value $\mathbb{S}_\alpha$.

In the absence of degeneracies, the singular subspace corresponding to 
$\mathbb{S}_\alpha$ is one--dimensional and therefore the symmetry operation must act 
as a proportionality transformation,
\begin{eqnarray}\label{eeqs}
\mathcal{Y}|\mathbb{V}_\alpha\rangle=\eta_\alpha\,|\mathbb{V}_\alpha\rangle, \nonumber \\
\mathcal{Y}|\mathbb{U}_\alpha\rangle=\eta_\alpha\,|\mathbb{U}_\alpha\rangle,
\end{eqnarray}
with $\eta_\alpha=\pm1$, since $\mathcal{Y}^2=\mathbb{1}$.

Thus, writing the singular vectors in Nambu form,
\begin{eqnarray}
|\mathbb{V}_\alpha\rangle=
\begin{pmatrix}
|\VV_\alpha\rangle\\
|\bar{\VV}_\alpha\rangle
\end{pmatrix},
\qquad
|\mathbb{U}_\alpha\rangle=
\begin{pmatrix}
|\UU_\alpha\rangle\\
|\bar{\UU}_\alpha\rangle
\end{pmatrix},
\end{eqnarray}
the eigenvalue Eqs. (\ref{eeqs})
imposes the following constraints
\begin{eqnarray}
\bar{\VV}_{j\alpha}=\pm i\,\VV_{j\alpha},\nonumber\\
\bar{\UU}_{j\alpha}=\pm i\,\UU_{j\alpha}.
\end{eqnarray}

\section{Effective incoherent pumping via adiabatic elimination}\label{app:anexoA}
In this section, we derive the effective model that arises when one eliminates adiabatically the fast decaying odd sites of a one-dimensional parametric chain by applying perturbation theory in Liouville space.

We consider two modes in the chain according to Fig. \ref{schesmes}. The first modes $a$ related to even sites decay with a rate given by $\gamma$, while the fast decaying odd sites decay with $\gamma'$, with $\gamma' \gg \gamma$. The master equation is given by
\begin{equation}
    \frac{d{\rho}}{dt} =\mathcal{L}_{\rm I}(\rho) + \mathcal{L}_{\gamma_{\rm d}}({\rho}),
\end{equation}
which contains the parametric interaction between the two modes,
$\mathcal{L}_{\rm I}(\rho) = - i \langle[ {H_{\rm I}}, \rho \rangle]$,  
with 
\begin{equation}
H_{\rm I} = g_{\rm c}'( b_{j+1} a_j +  b^\dagger_{j+1} a^\dagger_j+b_{j-1} a_j +  b^\dagger_{j-1} a^\dagger_j) ,
\nonumber
\end{equation}
and the radiative decay term
\begin{equation}
\mathcal{L}_{\gamma'} (\rho) 
= \frac{\gamma'}{2}(2 b_j \rho b^\dagger_j - b^\dagger_jb_j \rho - \rho b^\dagger_j b_j)
\nonumber
\end{equation}. 

We consider the limit $\gamma' \gg g_c'$, such that $\mathcal{L}_{\rm I}$ can be considered a small perturbation to the fast bosonic decay. 

Our aim is to adiabatically eliminate the fast decay terms and derive effective incoherent pumping. For this, we define the projection superoperator in Liouville space
\begin{equation}
    \mathcal{P}\rho=\rho_a\otimes \ket{0}\bra{0}_b,
\end{equation}
which projects into a product state of the reduced density matrix of the bosonic modes times the steady state of the auxiliary mode. Defining also the complementary projection operator $\mathcal{Q}=\mathds{1}-\mathcal{P}$, we look for a perturbative equation for the dynamics of $\mathcal{P}\rho$,
\begin{eqnarray}
    \mathcal{P}\dot{\rho} &= &\mathcal{P}\mathcal{L}_I\mathcal{Q}\rho, \\
    \mathcal{Q}\dot{\rho} &= &\mathcal{Q}\mathcal{L}_{\gamma'}\mathcal{Q}\rho+\mathcal{Q}\mathcal{L}_I\mathcal{P}\rho+ \mathcal{Q}\mathcal{L}_I\mathcal{Q}\rho,
\end{eqnarray}
where we have used that $\mathcal{L}_{\gamma'}\mathcal{P}\rho=0$ and $\mathcal{P}\mathcal{L}_I\mathcal{P}=0$. Taking use of the Nakajima-Zwanzig equation up to second order in the perturbative term, we can write,
\begin{equation}
    \mathcal{P}\dot{\rho}=\int^t_0 d\tau \mathcal{P}\mathcal{L}_I\mathcal{Q}e^{\mathcal{L}_{\gamma'}\tau}\mathcal{Q}\mathcal{L}_I\mathcal{P}\rho.
\end{equation}
Since we are interested in the dynamics of the bosonic modes $\dot{\rho}_a$, we should do the partial trace over the auxiliary degrees of freedom,
\begin{equation}\label{rdm}
    \dot{\rho}_a = \Tr_b{\int^t_0 d\tau \mathcal{P}\mathcal{L}_I\mathcal{Q}e^{\mathcal{L}_{\gamma'}\tau}\mathcal{Q}\mathcal{L}_I\mathcal{P}\rho}.
\end{equation}
We apply the definition of $\mathcal{P}\rho$ and $\mathcal{L}_I$ in terms of the commutators and develop the expression inside the integral, giving rise to
\begin{equation}\label{rhs2}
    \dot{\rho}_a = -\Tr_b \int^t_0 d\tau  
    \left[ H_{\rm I}, 
    \mathcal{Q}e^{\mathcal{L}_{\gamma'}\tau}\mathcal{Q}
    \left[
    H_{\rm I},\rho_a\otimes \ket{0}\bra{0}_b \right] \right] .
\end{equation}
To evaluate this expression, we need to use that
$\mathcal{Q}b^\dagger_j=b^\dagger_j$, $\mathcal{Q}b_j=b_j$, $\mathcal{L}_{\gamma'}(b^\dagger_j) = -
\frac{{\gamma'}}{2}b^\dagger_j$ and $\mathcal{L}_{\gamma'}(b_j) = -
\frac{{\gamma'}}{2}b_j$, and calculate all the terms that result from the double commutator in Eq. \eqref{rhs2}. Furthermore, we take the limit $t \to \infty$, which is well justified in the case of fast decay and we arrive at
\begin{equation}
    \dot{\rho}_a=P_{j,l}(2a^\dagger_j\rho_a a_l -a_ja^\dagger_l \rho_a - \rho_a  a_ja^\dagger_l),
\end{equation}
with the incoherent pumping matrix as
\begin{equation}
P_{j,l}=\frac{2g_c '^2}{\gamma'}(2\delta_{j,l}+\delta_{j,l+2}+\delta_{j,l-2})
\end{equation}
Note that the nonlocal pumping terms correspond to couplings to second-nearest neighbors in the full chain, which map onto nearest-neighbor interactions in the effective main chain. In addition, an extra local pumping term appears at each site of the chain.

\section{Renormalized Green's function in presence of weak  disorder}\label{Renorm}
We initiate our analysis with the Dyson equation, which is, in principle, valid for both Hermitian and non-Hermitian systems as shown in \cite{moustaj2022field}. 
Our goal is to find an approximation for the disordered averaged Green's function, 
$\langle \mathbb{G(\omega)}\rangle_{\rm dis}$, which we express in terms of the self-energy matrix, $\mathbb{M}$,
\begin{equation}
\langle \mathbb{G}(\omega) \rangle_{\rm dis}
= \frac{1}{\mathbb{G}_0(\omega)^{-1} - \mathbb{M}},
\end{equation}
This equation can be re-written as 
\begin{equation}
\langle \mathbb{G}(\omega) \rangle_{\rm dis}
= \mathbb{G}_0(\omega) 
+ \mathbb{G}_0(\omega) \mathbb{M} 
\langle \mathbb{G}(\omega) \rangle_{\rm dis}.
\end{equation}
We define the disorder Hamiltonian term
\begin{equation}
    \mathbb{W}
    =\begin{pmatrix}
        {\rm diag}({\bf w}) & 0 \\ 
        0 & -{\rm diag}({\bf w})
    \end{pmatrix},
\end{equation}
where $w_j$ are independent Gaussian variables with zero mean and standard deviation $W$ representing random values of the local bosonic mode energies. The disorder averaged Green's function can be expressed like
\begin{eqnarray}
\langle \mathbb{G} (\omega)\rangle_{\rm dis} 
= \langle \frac{1}{\mathbb{G}_0(\omega)^{-1} - \mathbb{W}} \rangle_{\rm dis} 
\end{eqnarray}
We can write the Dyson series
\begin{align}\label{Dyson}
    \langle \mathbb{G}(\omega) \rangle_{\rm dis} 
    & = \mathbb{G}_0(\omega) + 
    \langle \mathbb{G}_0(\omega) \mathbb{W} \mathbb{G}_0(\omega) \rangle_{\rm dis}
\nonumber \\
    & +
    \langle \mathbb{G}_0(\omega) \mathbb{W} \mathbb{G}_0(\omega)
    \mathbb{W} \mathbb{G}_0(\omega) \rangle_{\rm dis} + 
\nonumber \\
    & +
    \langle \mathbb{G}_0(\omega) \mathbb{W} \mathbb{G}_0(\omega)
    \mathbb{W} \mathbb{G}_0(\omega) \mathbb{W}  \mathbb{G}_0(\omega) \rangle_{\rm dis} + \dots
\end{align}
In the last equation, since $\langle \mathbb{W} \rangle_{\rm dis} = 0$, odd powers of the disorder matrix cancel. In the Born-approximation, we approximate
\begin{align}
& \langle \mathbb{W} \mathbb{G}_0(\omega)
    \mathbb{W} \dots
    \mathbb{W}  \mathbb{G}_0(\omega) \mathbb{W} 
    \rangle_{\rm dis} \approx \langle \mathbb{W} \mathbb{G}_0(\omega)
    \mathbb{W} \rangle_{\rm dis} \dots
\langle \mathbb{W}  \mathbb{G}_0(\omega) \mathbb{W} 
\rangle_{\rm dis} ,
\end{align}
which is equivalent to the following approximation for the self-energy.
\begin{equation}
\mathbb{M} \approx \mathbb{M}_{\rm BA} =
\langle \mathbb{W} \mathbb{G}_0(\omega) \mathbb{W}  \rangle_{\rm dis} .
\end{equation}
This quantity can be explicitly calculated, yielding, in terms of the Green's function blocks defined in Eq. \eqref{eq:G.blocks},
\begin{equation}
\mathbb{M}_{\rm BA}
= W^2 
\begin{pmatrix}
{\rm diag}(G) & -{\rm diag}(\bar{G}) \\
        -{\rm diag}(\bar{G}') & {\rm diag}(G')
    \end{pmatrix},
\end{equation}
From this equation, one can infer the effective parameters in the topological phase within the symmetric parameter regime ($\Delta=0$, $\phi=\pi/2$)
\begin{eqnarray}\label{ren}
    \Delta_{\rm eff}&\approx& \Delta + W^2\Re{(G_{ii})}, \nonumber \\
    \gamma_{\rm eff}&\approx&\gamma -2 W^2\Im{(G_{ii})}, \nonumber \\
    g_{s_{\rm eff}}&\approx& g_s-W^2(\Bar{G}_{ii}).
\end{eqnarray}
Furthermore, from Eqs. (\ref{V0}, \ref{U0}), one verifies that the effective parameters in the topological phase are spatially homogeneous. In particular, $G_{ii}\approx\frac{1}{4}e^{-i\phi_U}e^{\lambda}$, while $\bar{G}_{ii}$ differs from $G_{ii}$ only by an additional factor of $i$.

\section{Analytical expressions for zero singular vectors and singular values }\label{sv}
In this section, we derive analytical expressions for topological zero singular values and edge vectors in the large chain limit $N \gg 1$. The algebraic method proceeds as follows: (i) we map the singular value decomposition of the dynamical matrix onto the eigendecomposition of an enlarged Hermitian matrix. (ii) Assuming that the hybridization between left- and right-localized singular vectors is negligible for $N \gg 1$, we approximate the smallest singular value as $\mathbb{S}_0\approx0$, which enables us to obtain closed-form expressions for the corresponding singular vectors. (iii) We relax this approximation and use the resulting analytical expressions to compute the finite zero value of $\mathbb{S}_0$ for finite system sizes.

\subsection{Calculation of edge singular vectors}
An analytical method for calculating left and right edge singular vectors, $|\mathbb{U}_0\rangle$, $|\mathbb{V}_0\rangle$, can be derived in the limit 
$N \gg 1$. Finite-size corrections will be addressed below. In this limit, the associated zero-singular value vanishes asymptotically, $\mathbb{S}_0 \approx 0$, such that, 
\begin{eqnarray}\label{aprox}
(\omega - \mathbb{H}) |\mathbb{V}_0 \rangle \approx 0 , \qquad 
(\omega - \mathbb{H}^\dagger) |\mathbb{U}_0 \rangle  \approx 0 .
\end{eqnarray}
For the sake of simplicity, we restrict our analysis to the symmetric parameter regime defined by $\Delta = 0$, and $\phi = \pi/2$, which is assumed throughout this work. 
$\mathbb{H}$ is invariant under symmetry $\mathcal{Y}$ defined in App. \ref{SPR}. 
Thus, the $|\mathbb{V_0} \rangle$ can be written as eigenstates of $\mathcal{Y}$,
\begin{eqnarray}
|\mathbb{V}^{\pm}_0 \rangle=
\begin{pmatrix}
|\VV_0^{\pm} \rangle\\
\pm i |\VV_0^{\pm} \rangle
\end{pmatrix},
\label{rel}
\end{eqnarray}
Replacing Eq. (\ref{rel}) into (\ref{aprox}),  we obtain two independent homogeneous second-order difference equations as
\begin{align}
(\omega  + i\frac{\gamma}{2}\pm ig_s)  \VV^{\pm}_{j0} +  
     i (J\pm g_c)\VV^{\pm}_{j+1,0}-i(J\mp g_c)\VV^{\pm}_{j-1,0} \approx0,
\end{align}
where each equation is associated with an independent parity sector $(\pm)$. 
We use an exponential ansatz of the form $V_{0j} \propto \beta^j$, an approach 
that is equivalent to the non-Bloch formalism \cite{yao2018edge,yokomizo2019non}, and  get the characteristic equation 
\begin{equation}
i(J\pm g_c)\beta^2+(\omega+i\frac{\gamma}{2}\pm ig_s)\beta-i(J\mp g_c)\approx0. 
\end{equation}
Each $(\pm)$--channel admits two solutions $\beta_{1,2}$ such that the general solution can be written as a linear combination
\begin{eqnarray}
    \VV_{j0}^{\pm}=C^{\pm}_1 \beta_{1\pm}^{j}+C_2^{\pm} \beta_{2\pm}^j,
\end{eqnarray}
where each $\beta_{i\pm}$ is a complex number associated with one of the two independent channels. A mode is localized at the left (right) boundary when 
$|\beta_i|<1$ ($|\beta_i|>1$). 
The coefficients $C_i^{\pm}$ are fixed by considering extra sites at $j = 0$ and $j= N + 1$, and 
imposing boundary conditions on those extra sites:
\begin{eqnarray}
    \VV_{j = 0,0}=0,\\
    \VV_{j = N+1, 0}=0.
\end{eqnarray}
As a side remark, we notice that the order of the characteristic equation is determined by the number of left- and right-hopping terms. In the present case, the recurrence relation involves one hopping term to the left, $\VV_{j-1}$, and one to the right, $\VV_{j+1}$, yielding two independent modes. 
In the most general situation, if we had $L$ hoppings two the left and $R$ hoppings to the right, the solution would contain $L+R$ independent modes,
\begin{eqnarray}
\VV_{j0} =\sum_s^{L+R}C_s\beta_s^j,
\end{eqnarray}
and the same number of boundary conditions must be imposed in order to fix these coefficients.

\subsubsection{Analytical solution in the limit of semi-infinite chains}
A particularly relevant case occurs when $J =g_s= g_c = g$, which is commonly considered throughout this work (see Sec. \ref{modelI}). In this regime, the system reduces to two independent first-order linear difference equations,
\begin{eqnarray}   
(\omega+i\frac{\gamma}{2}+ g_s)\VV^{-}_{j0}
+
2ig\VV^{-}_{j+1,0}\approx0, 
\label{eq:primera}
\\
(\omega+i\frac{\gamma}{2}- g_s) \VV^{+}_{j0} - 2 i g\VV^{+}_{j-1,0}\approx0 \label{eq:segunda},
\end{eqnarray}
Since each equation involves only a single hopping term, the general solution consists of a single mode rather than a linear combination, i.e.,
\begin{eqnarray}\label{ansatz}
\VV_{j0}^{\pm} = A^{\pm} \beta_{\pm}^{j},
\end{eqnarray}
where $A^{\pm}$ are normalization constants. 
Because only one coefficient appears in each solution, a single boundary condition suffices. By imposing the boundary condition at one edge, we effectively treat the system as a semi-infinite chain. 
Specifically, the $(-)$ channel admits a boundary condition on the left edge, while the $(+)$ channel admits a boundary condition on the right edge.

\begin{figure}[h]
    \centering
    \includegraphics[width=0.65\textwidth]{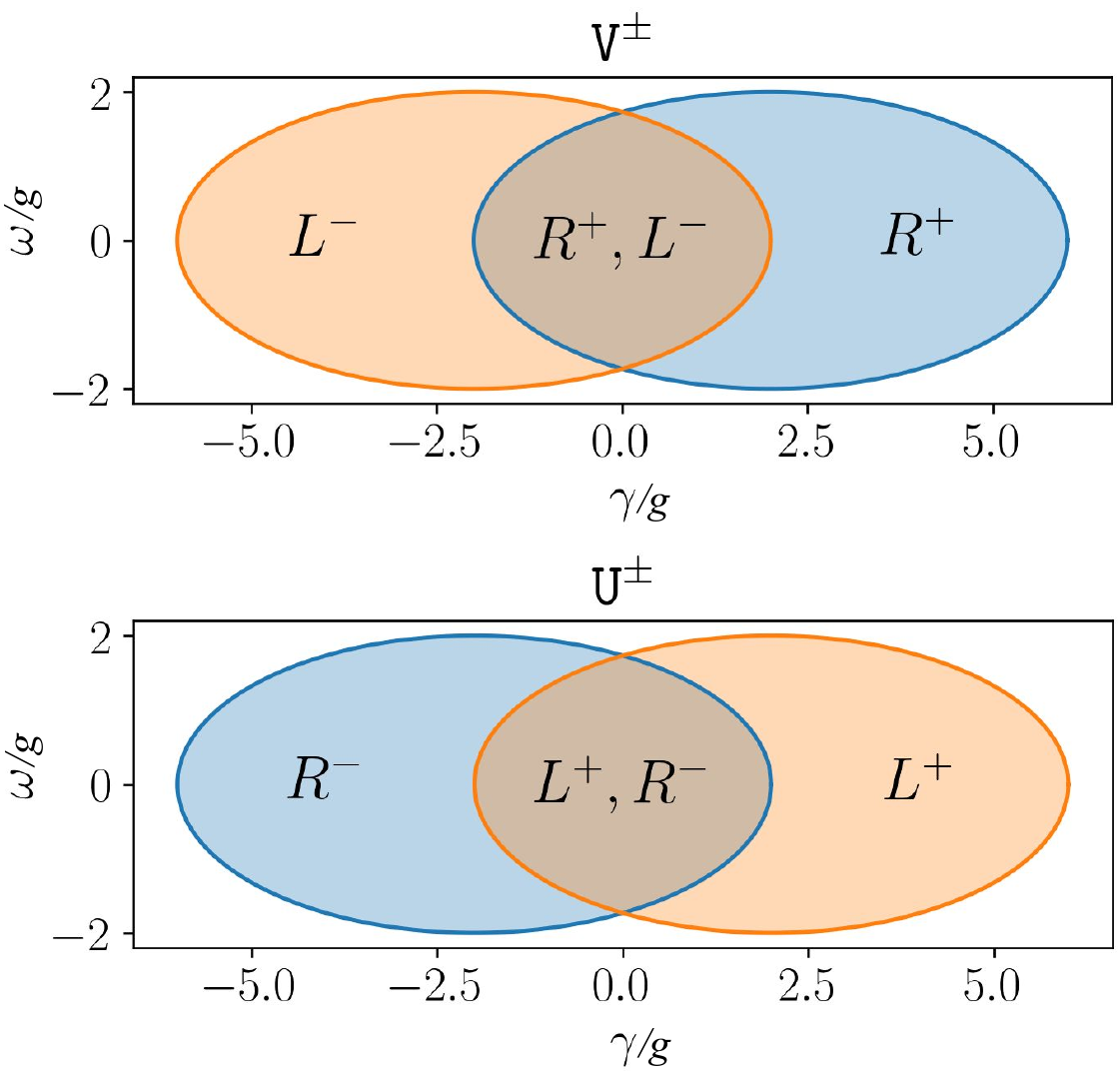}
    \caption{Edge-state solutions for $\VV^{\pm}$ (top) and $\UU^{\pm}$ (bottom) lying within the boundaries defined by the ellipses in Eq.~(\ref{elipseq}). Orange and blue shading indicate states localized on the left and right edges, respectively. The remaining parameters correspond to the symmetric parameter regime defined in Sec. \ref{modelI}.}
    \label{elipses}
\end{figure}

After replacing Eq. (\ref{ansatz}) into Eqs. (\ref{eq:primera}) and (\ref{eq:segunda}), 
we get the following independent solutions
\begin{eqnarray}
\beta_{-} &=& - \frac{\gamma + 2g}{4g}+i\frac{\omega}{2g}, \\
\beta_{+}^{-1} &=&  \frac{\gamma-2g}{4g}-i\frac{\omega}{2g}. 
\end{eqnarray}
The left localized  $(-)$ channel solution is normalizable if $|\beta_{-}| < 1$, while the right localized $(+)$ channel solution is normalizable if $|\beta_{+}|>1$.
This condition immediately yields
\begin{eqnarray}\label{elipseq}
\frac{( \gamma \pm 2g)^2 + 4\omega^2}{16 g^2} \leq 1,
\end{eqnarray}
which defines the regions in parameter space 
$(\gamma,\omega)$ where left- and/or right-localized edge modes exist. From Eq.~(\ref{elipseq}), it follows that the above inequality is satisfied whenever the parameters lie on or within each of the ellipses shown in Fig.~\ref{elipses}. Since dissipation in this model is a positive quantity $\gamma/g>0$, it gives us three different regions for $\omega/g=0$:
\begin{itemize}
    \item $0<\gamma/g<2$: Edge states $\VV_{0j}$ ($\UU$) are localized at both edges simultaneously (trivial amplifying configuration).
    \item $2<\gamma/g<6$: A single edge vector $\VV$ $(\UU)$ is localized at the right (left) edge (topologically protected amplifying regime).
    \item $\gamma/g>6$: No edge localization occurs (trivial non-amplifying configuration).
\end{itemize}
These results are consistent with previous studies \cite{gomez2023driven}.

We define $\beta\equiv e^{iq}$, with $q\in \mathbb{C}$, with $q\equiv\Tilde{k}-i\lambda$, where $\Tilde{k}$ represents the generalized Bloch momentum, while $\lambda$ is the inverse localization length giving rise to $|\beta|\neq 0$.
From the previous solutions, we obtain  
\begin{eqnarray}
    \lambda^{-}(\omega)&=&+\frac{1}{2}\ln\!\left[\frac{(\gamma+2g)^2+4\omega^2}{16g^2}\right],\\
    \lambda^{+}(\omega)&=&-\frac{1}{2}\ln\!\left[\frac{(\gamma-2g)^2+4\omega^2}{16g^2}\right].
\end{eqnarray}
Equivalently, the complex phases for each solution are
\begin{eqnarray}
\Tilde{k}^{-}(\omega) & =& -\arctan\!\left(\frac{2\omega}{\gamma+2g}\right), \\
\Tilde{k}^{+}(\omega) &= & \arctan\!\left(\frac{2\omega}{\gamma-2g}\right).
\end{eqnarray}

\begin{figure}[h]
    \centering
    \includegraphics[width=0.75\textwidth]{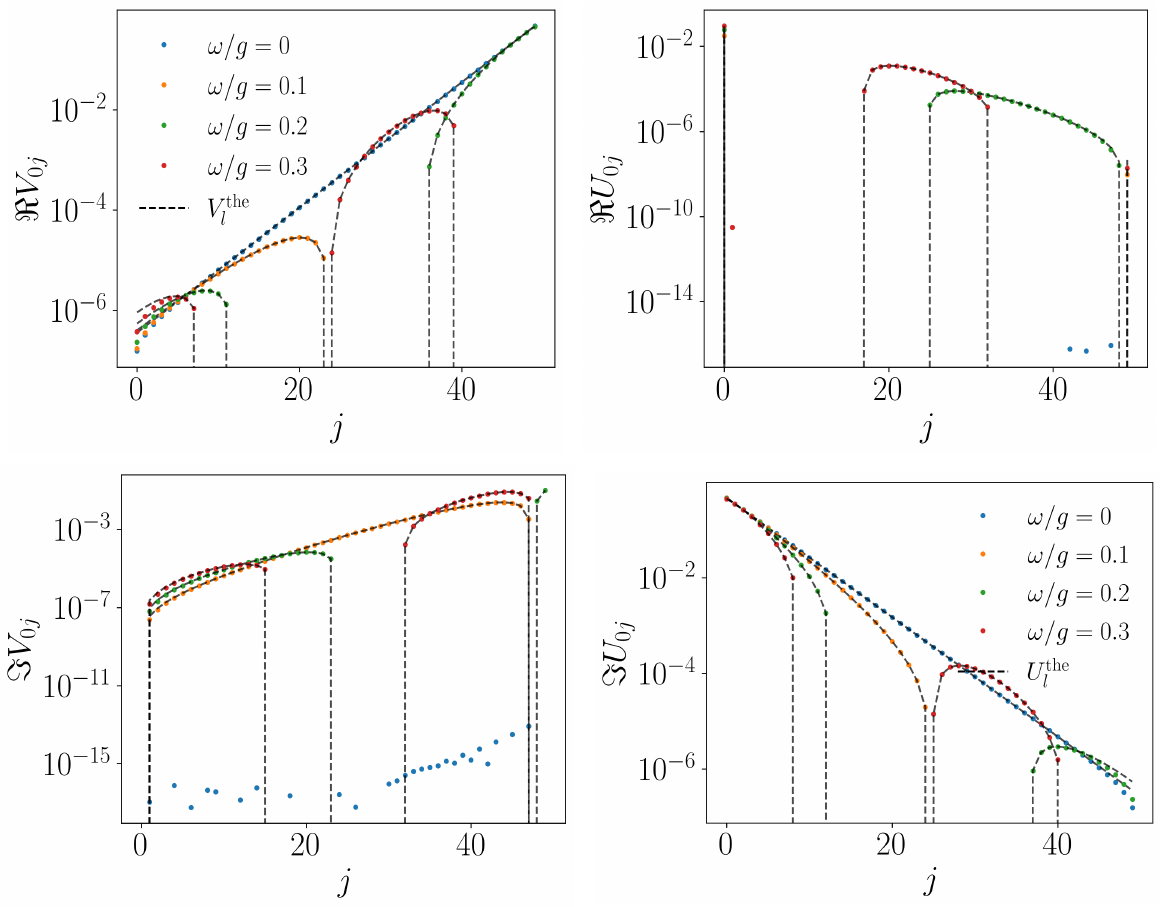}
    \caption{Comparison between numerical values (dots) and analytical expressions (dashed lines) of the real and imaginary part of the singular vectors $\VV_{l0}$ and $\UU_{l0}$ for different values of the frequency $\omega/g$ and $\gamma/g=5$. The remaining parameters correspond to the symmetric parameter regime defined in Sec. \ref{modelI}.}
    \label{UVNR}
\end{figure}

We repeat the calculation for the left singular vector, which must satisfy 
$(\omega - \mathbb{H}^\dagger) |\mathbb{U}_0\rangle = 0$. 
The final form for both singular vectors is then (see Fig. \ref{UVNR})
\begin{eqnarray}
\VV_{l0} &=& A \, e^{(i\tilde{k} + \lambda)l}, \label{V0}\\
\UU_{l0} &=& e^{i \phi_U} A \,  
e^{i\tilde{k} l} e^{\lambda(N-1-l)} \label{U0}.
\end{eqnarray}
We have written the solutions such that they have a common normalization factor, $A$, and omitted the subscript $^{\pm}$ in $A$, $\lambda$ and $k$. The phase factor $e^{i \phi_U}$ indicates that, at this point of the calculation, the phase difference between $\VV_{l0}$ and $\UU_{l0}$ is a free parameter. 
However, this phase is fixed once we require those singular vectors to form an eigenvector of the extended matrix $\cal H (\omega)$ in \eqref{HUS}. We prove in the next section that this condition sets 
\begin{equation}
\phi_U = \frac{\pi}{2} - \tilde{k}.
\label{phiU}
\end{equation}

By imposing the standard singular vector normalization $\sum_l\VV_{l0}^*\VV_{l0}+\bar{\VV}_{l0}^*\bar{\VV}_{l0}=1$, the coefficient $A$ is given by
\begin{eqnarray}
A &=& \frac{1}{\sqrt{2}}
\sqrt{\frac{e^{2\lambda} - 1}{e^{2\lambda N} - 1}}.
\end{eqnarray}
The analytical expressions fits perfectly numerical calculations as shown in Fig. \ref{UVNR}.

\subsection{Finite-size corrections: Singular values}
So far we have imposed semi-infinite boundary conditions in the limit $N \gg 1$, in which $\mathbb{S}_0 = 0$. The singular vectors $| \mathbb{U}_0 \rangle$ and $| \mathbb{V}_0 \rangle$ will form a topological edge state of the hermitized Hamiltonian ${\cal H}(\omega)$. This observation allows us to calculate finite size corrections to $\mathbb{S}_0$ from the hybridization between those vectors. 

From Eq. (\ref{HUS}), the zero singular value can be obtained as:
\begin{eqnarray}
    \mathbb{S}_0=\sum_{\mu,\nu=0}^{2N-1}\mathbb{U}^*_{\mu0}(\omega\mathbb{1}_{2N}-\mathbb{H})_{\mu\nu}\mathbb{V}_{\nu0}.
\end{eqnarray}
We focus on the topologically protected regime with a single normalizable solution in the positive parity channel with 
$\Tilde{\VV}_{l0}=+i\VV_{l0}$, and $\Tilde{\UU}_{l0}=+i\UU_{l0}$. After some algebra, we obtain
\begin{equation}
\mathbb{S}_0 =2 \sum_{j,l = 0}^{N-1} 
\UU_{j0}^*
\left(\omega\mathbb{1}-\msf{J} + 
i\frac{\Gamma}{2}-i \msf{K} \right)_{jl}
\VV_{l0}.
\end{equation}
Using the expression above this quantity can be rewritten as
\begin{eqnarray}
\mathbb{S}_0 =
2
\left(
\omega+i\frac{\gamma-g}{2}\right) \langle \UU_0|\VV_0 \rangle
-2\langle \UU_0 | \msf{J} + 
i \msf{K}^{\rm off\text{-}diag}| \VV_0 \rangle, \nonumber
\end{eqnarray}
where $\msf{K}^{\rm off-diag}$ is the parametric coupling $\msf{K}$ with only off-diagonal terms (proportional to $g_c$).

The overlap between the left and right singular vectors can be calculated as
\begin{equation}
\langle \UU_0 |\VV_0 \rangle 
\equiv \sum_{l=0}^{N-1} \UU_{l0}^*\, \VV_{l0} = 
-\frac{i}{2} e^{i\tilde{k}} N
\frac{(1-e^{2\lambda})}
{(1-e^{2\lambda N})}e^{\lambda(N-1)}.
\end{equation}

The second contribution can be evaluated explicitly, yielding
\begin{align}
\langle \UU_0 |
\msf{J} &+ i \msf{K}^{\rm off\text{-}diag}|\VV_0 \rangle \equiv 
\sum_{j,l=0}^{N-1}
\UU_{j0}^*(\msf{J}  + i \msf{K}^{\rm off\text{-}diag})_{jl} \VV_{l0}
\nonumber \\ &= 
2i\sum_{l=0}^{N-1}\UU_l^*\VV_{l-1}
=
(N-1)e^{-\lambda}
\frac{(1-e^{2\lambda})}
{(1-e^{2\lambda N})}e^{\lambda(N-1)}.
\end{align}
Combining both terms, and using the definition of $\beta_{+}^{-1} $ as $e^{-i\tilde{k}-\lambda} = \frac{\gamma-2}{4} - i\frac{\omega}{2}$, we obtain
\begin{eqnarray}
\mathbb{S}_0 =2e^{-\lambda}\frac{(1-e^{2\lambda})}
{(1-e^{2\lambda N})}e^{\lambda(N-1)}.
\end{eqnarray}
In the thermodynamic limit $N\to\infty$, the singular value exhibits the following exponential scaling,
\begin{equation}
\mathbb{S}_0 \approx
2(1-e^{-2\lambda })e^{-\lambda N}.
\end{equation}

\begin{figure}[h]
\centering
\begin{minipage}{0.4\textwidth}
    \centering
    \includegraphics[width=\linewidth]{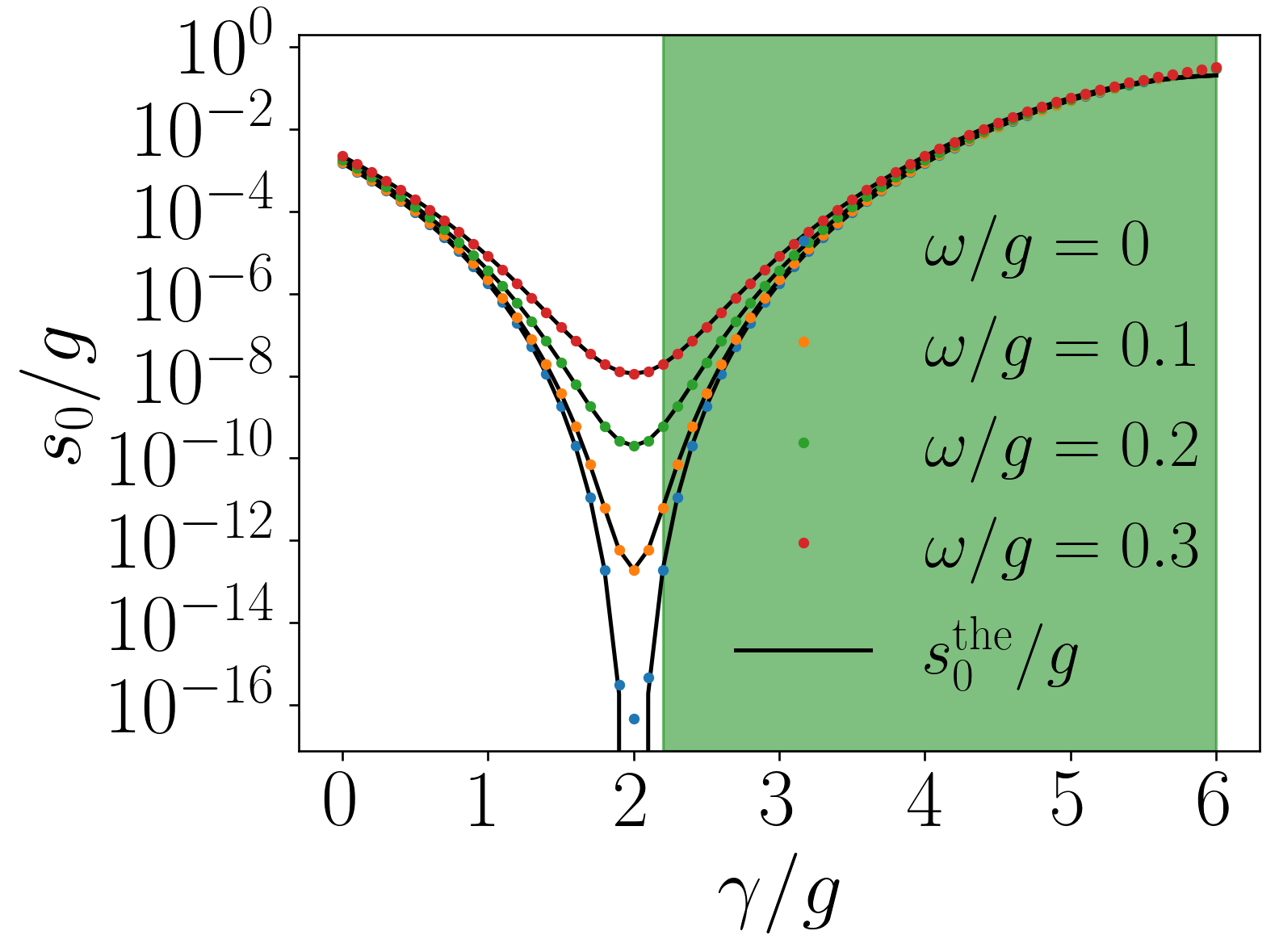}
    \caption{Numerical values (dots) and analytical expressions (solid black lines) of the zero singular values for different values of the frequency $\omega/g$. The green shaded region corresponds to the stable regime for $N=10$. Parameter set defined in Sec. \ref{modelI}, corresponding to Model I.}{}
    \label{s0}
\end{minipage}
\hfill
\begin{minipage}{0.4\textwidth}
    \centering
    \includegraphics[width=\linewidth]{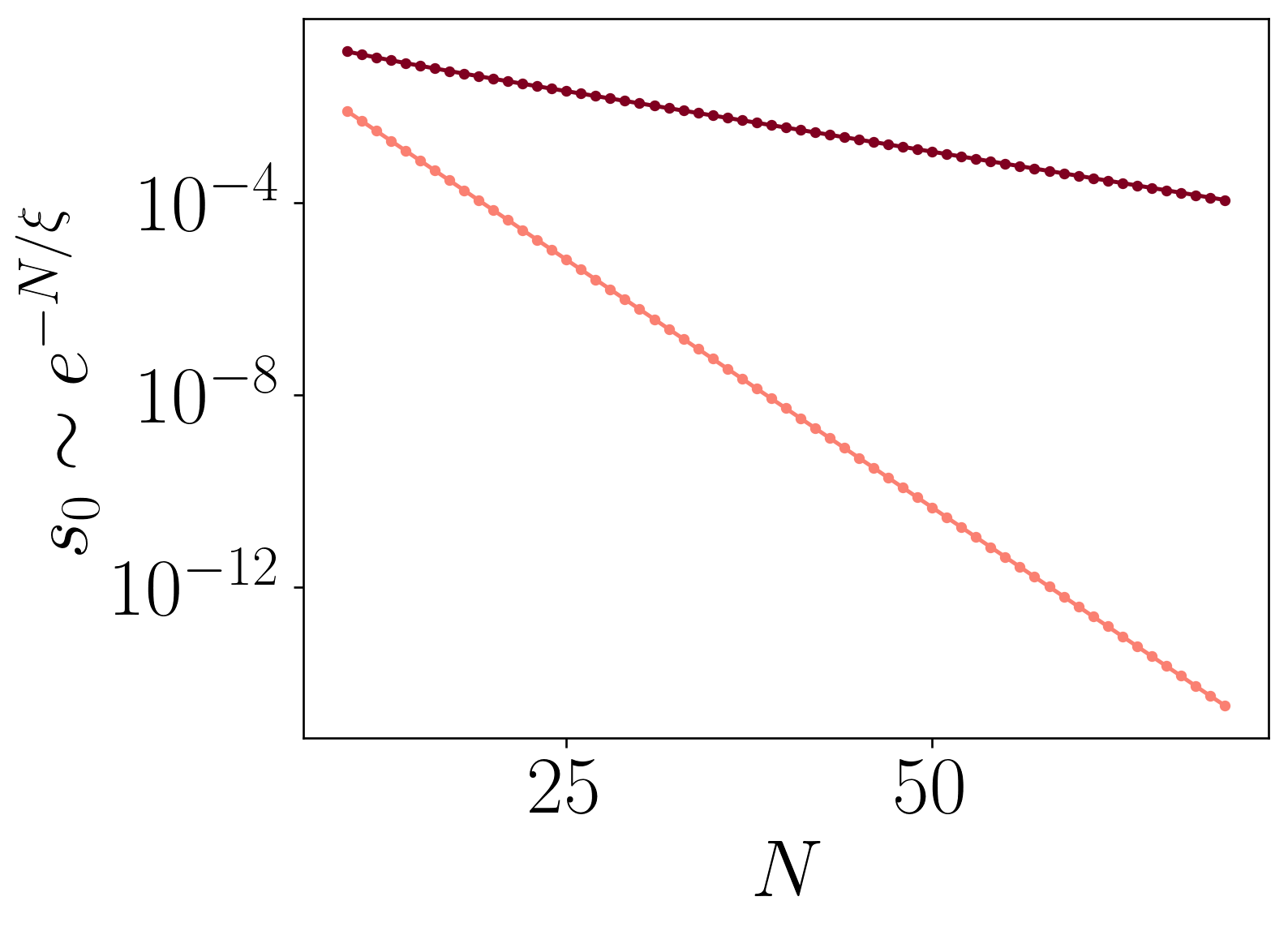}
    \caption{Zero singular value in the topological phase with winding number $\nu(\omega)=2$ as a function of the system size $N$. Parameters are given by the set defined in Sec. \ref{modelII}, with $\gamma/g=3.75$, $g_c'/g=3$, $\gamma'/g=30$ (corresponding to $P/g=\frac{3}{5}$), and $\omega/g=0$.}{}
    \label{s1s2}
\end{minipage}
\end{figure}

This discussion has been carried out for model I with $P=0$, and the analytical expression fits numerical calculation as illustrated in Fig. \ref{s0}. However, higher values of $P$ can lead to $\nu>1$, which leads to many topologically protected zero modes. In practice, as we can check in Fig. \ref{s1s2}, one of the two zero singular values is more suppressed that the other in the thermodynamic limit. Thus, most of the results regarding LRO also apply for this case of many topologically protected zero modes.

\end{appendices}

\end{document}